\documentclass[12pt,a4paper,final]{iopart}

  \expandafter\let\csname equation*\endcsname\relax
  \expandafter\let\csname endequation*\endcsname\relax
  
\usepackage[utf8]{inputenc}
\usepackage[english]{babel}
\usepackage{graphicx}
\usepackage{amsmath}
\usepackage{amssymb}
\usepackage{color}
\usepackage{bbold}
\usepackage[abs]{overpic}
\usepackage{mathtools}

\usepackage{cancel}

\newcommand{\Id}{{\mathbb 1}}

\newcommand{\eN}{\mathcal{N}}

\newcommand{\baN}{\bar{N}}

\newcommand{\ie}{i.e.~}
\newcommand{\eg}{e.g.~}

\definecolor{darkgreen}{rgb}{0,0.5,0}

\begin{document}

\title{Lattice Gauge Tensor Networks}

\author{Pietro Silvi$^{1}$, Enrique Rico$^{1,2}$, Tommaso Calarco$^{1,3}$, Simone Montangero$^{1,3}$}
\address{$^1$Institut f\"ur Quanteninformationsverarbeitung, Universit\"at Ulm, D-89069 Ulm, Germany}
\address{$^2$IPCMS (UMR 7504) and ISIS (UMR 7006), Universit\'e de Strasbourg and CNRS, Strasbourg, France}
\address{$^3$Center for Integrated Quantum Science and Technology (IQST), Universities of Ulm/Stuttgart and MPI for Solid State Research}
\ead{pietro.silvi@uni-ulm.de}

\date{\today}

\begin{abstract}
We present a unified framework to describe lattice gauge theories by means of tensor networks: this framework is efficient as it exploits the high amount of local symmetry content native of these systems describing only the gauge invariant subspace. Compared to a standard tensor network description, the gauge invariant one allows to speed-up real and imaginary time evolution of a factor that is up to the square of the dimension of the link variable. The gauge invariant tensor network description is based on the quantum link formulation, a compact and intuitive formulation for gauge theories on the lattice, and it is
alternative to and can be combined with the global symmetric tensor network description. We present some paradigmatic examples that show how this architecture might be used to describe the physics of condensed matter and high-energy physics systems. Finally, we present a cellular automata analysis which estimates the gauge invariant Hilbert space dimension as a function of the number of lattice sites and that might guide the search for effective simplified models of complex theories. 
\end{abstract}


\maketitle

\section{Introduction}
In modern science, gauge theories play a fundamental role, with examples ranging from quantum electrodynamics to the standard model of elementary particle physics~\cite{LGT1,LGT2}. They represent a cornerstone in our understanding of the physical world and lie at the heart of theories dealing with such diverse systems as quantum spin liquids and the quark-gluon plasma. Starting from the breakthrough contribution by Wilson in 1974~\cite{Wilson,Kogut}, Lattice Gauge Theories (LGT) have attracted a significant attention across several branches of theoretical physics. While the lattice formulation of gauge theories has intrinsic fundamental interest in high-energy physics due to the prominent role played by gauge fields, emergent gauge models have also been introduced in different condensed matter setups in relation to exotic many-body phenomena such as quantum spin liquids and topological states of matter~\cite{LNW,LMM,cond}. Furthermore, there is good reason to believe that certain types of gauge structures could open up new possibilities for quantum computation~\cite{Kitaev,Nayak}. 

Quantum simulation of gauge theories is receiving an increasing degree of interest~\cite{NatureInsightReview}, due to the fact that these types of platforms could provide the tools to simulate dynamical properties and/or out of equilibrium physics in lattice gauge models including fermionic matter fields, as they are sign-problem
free simulators by construction. In the context of atomic, molecular and optical physics, several proposals for the quantum simulation of lattice gauge theories have been made recently~\cite{Zohar,Banerjee,Tagliacozzo,Hauke,Marcos,Wiese2,zohar2013,stannigel2013,shimizu}.
These quantum analog simulations can be seen as a complementary tool to the existing classical ones --the latter could also benchmark the outcomes of the former.

The prediction power of quantum gauge theories is often limited due to the fact that the computational resources needed to perform simulations involving large numbers of particles are typically forbidding. The present manuscript presents a partial solution to this problem showing that the gauge invariant subspace of these theories can be represented exactly by a local set of tensor networks, thus it is possible to apply the well developed and successful architecture of tensor networks to lattice gauge theories. In other words, we present an exact and efficient tensor network representation of abelian and non-abelian lattice gauge theories~\cite{Enrique13}. 

Tensor networks are one of the mainstream paradigms for simulating quantum many-body lattice systems, both in and out of equilibrium, via a representation of the quantum state with tailored variational ansatz wavefunctions. They originated from the understanding that the density matrix renormalization group (DMRG) technique~\cite{White92} could be recast in a variational formulation, by means of matrix product states~\cite{MPS,DMRGMPS1,DMRGMPS2,PrimoMPS1,PrimoMPS2,PrimoMPS3,PrimoMPS4,PrimoMPS5}. This stimulated further development of such framework in the last decade in several directions extending the tensor network paradigm to encompass higher dimensionality~\cite{PrimoPEPS}, peculiar geometries~\cite{Dendrimer,PrimoMERA}, and the limit to the continuum~\cite{CMPS}. 

One of the most appealing features portrayed by tensor network is the possibility of encoding and controlling global symmetries of the local degrees of freedom~\cite{Singh1,Singh2}, that characterize several condensed matter models. In fact, a general, robust, and numerically efficient formulation of any of such symmetries in the Tensor Network framework is known~\cite{Singh3,Singh4}; it is commonly used in simulation to achieve an enhancement of the algorithm performance, as well as a precise targeting of irreducible representation sectors~\cite{Bauer,Weichselbaum,CorbozSU3}.

Lattice gauge symmetries differ from global ones, since they have quasi-local supports and are typically homogeneous, yielding a combined Lie algebra of generators which grows extensively with the system size. Nevertheless, several physical contexts have been found where tensor networks are an exact description of ground states of gauge-invariant Hamiltonians, e.g., 2D toric code that is an Ising gauge theory \cite{wegner,Kitaev,Ardonne}. More recently, this framework has been successfully applied to LGT related problems \cite{Martin,Byrnes,Haegeman,Luca,Banuls,Silvi,Enrique13,buyens}. In fact, Tensor Networks represent microscopically the local Hilbert spaces and at the same time are tailored on a real-space wave-function representation, so they can be used to describe real-space locality and local symmetries altogether.

Here we show how Tensor Network can exactly encode lattice gauge symmetries providing an architecture that is completely general and computationally efficient: our approach outperforms a straightforward approach that do not explicitly exploits gauge symmetries. To achieve this goal, the use of alternative formulations of gauge theories is highly desirable, the principal motivation being the identification of models with a finite dimensional Hilbert space at each link or site which can be simulated by tensor networks algorithms. Thus, we develop this architecture in the Quantum Link Model (QLM) formulation~\cite{Horn,Orland,Wiese} of Hamiltonian lattice gauge theories. Wilson's formulation of lattice gauge theory has an infinite dimensional Hilbert space at each link due to the use of continuously varying fields~\cite{Wilson}. Quantum link models provide a complementary formulation of lattice gauge theories introducing generalized quantum spins associated with the links of a lattice. In fact, under some physically motivated assumptions, Wilson's lattice gauge theories can be obtained from QLM~\cite{Brower,BCUJW04}.

One possible way to approach the continuum limit for quantum link models is by dimensional reduction from a higher dimension where continuous gluon fields then arise as collective excitations of discrete quantum link variables, in the same way as magnons arise as collective excitations of quantum spins~\cite{Wiese2}. Such extra dimension is expected to be exponentially smaller than the actual spatial dimension, thus not posing a threat to numerical methods. A second strategy is to increase the number of rishons per link, which also have been seen as a way to achieve the continuum limit without requiring the extra dimension~\cite{zohar2013} (we will investigate numerical feasibility of this strategy later on).

In addition, there are several examples of condensed matter models, characterized by lattice gauge symmetries, where the gauge degrees of freedom are inherently finite-dimensional. This is the case, for instance, for spin-ice or quantum dimer models~\cite{anti} or in discrete gauge models like the Ising gauge theory~\cite{wegner}.

The formulation of lattice gauge tensor network we present here in details, allows to represent efficiently and exactly the gauge constraints of this classes of systems, with a performance that improves up to quadratically with the quantum link dimension, and thus it increases its efficiency at the Wilson limit. 

The manuscript is organized as follows: In Sec.~\ref{sec:qlm} we review the framework to describe lattice gauge theories into quantum link formulation. In Sec.~\ref{sec:MPO} we provide a constructive scheme to embed the QLM picture within the Tensor Network framework, which relies on matrix product formalism in 1D (and projected entangled pair formalism in higher dimensions). The algorithm to exploit such representation in numerical context is described in Sec.~\ref{sec:speedup}, mainly focusing on time evolution (both in real and imaginary time). In Sec.~\ref{sec:automata} we perform some theoretical scaling investigation of effective Hilbert spaces growth, under the QLM constraints, made easily available through the tensor network picture. Finally, in Sec.~\ref{sec:conc} we draw our conclusions.

\section{Quantum Link Models} \label{sec:qlm}

From now on, as we focus on numerical simulations, we assume that the space of the gauge degrees of freedom is finite dimensional. Starting from this assumption, the formulation in quantum link model language of lattice gauge theories follows without additional loss of generality~\cite{Horn,Orland,Wiese}. We define the gauge invariant model of interests by defining three elements:
\begin{itemize}
\item The {\bf local degrees of freedom} [$\psi_x^{a}, U^{ab}_{x,x+\mu_x}, E_{x,x+\mu_x}$] - We describe as quantum degrees of freedom both the lattice sites, which we will refer to as `matter field', and the `gauge field' and its canonical conjugate variable or `electric field' located on the links (the lattice bonds between neighboring sites, every link being shared by a different pair of sites).
\item The {\bf gauge symmetry generators} [$G^{\nu}_x$] - unlike global symmetries, which operate nontrivially upon the whole lattice, gauge symmetry generators have a localized support, each one involving a single matter field site, and all the gauge fields connected to it.
\item The {\bf gauge invariant dynamics} [$H$] - The dynamics is defined via a Hamiltonian which commutes with the whole algebra of gauge generators, which guarantees that gauge invariance is conserved throughout the time evolution (as in Fig.~\ref{fig:ingredients}, panel a).
\end{itemize}

In this section, we analyze in detail these elements in a quantum link formulation, while stressing the connection to typical lattice gauge theory models.

\subsubsection{Local degrees of freedom.}

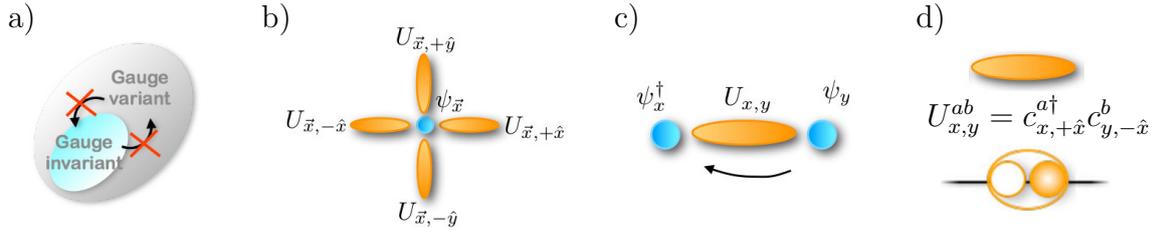
\begin{figure}
 \begin{center}
 \begin{overpic}[width = \columnwidth, unit=1pt]{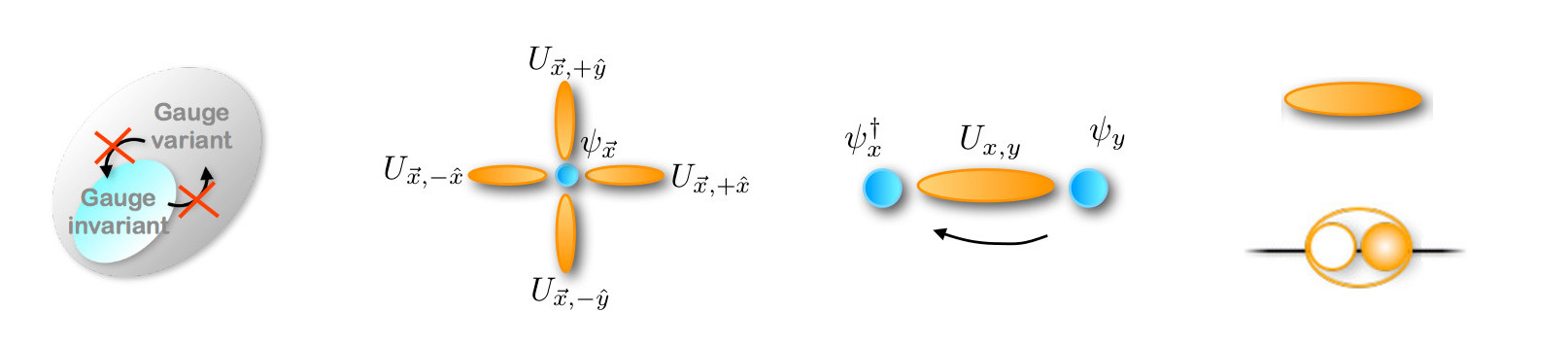}
  \put(5, 86){a)}
  \put(100, 86){b)}
  \put(232, 86){c)}
  \put(345, 86){d)}
  \put(349, 49){$U^{ab}_{x,y} = c^{a\dagger}_{x,+\hat{x}} c^{b}_{y,-\hat{x}}$} \end{overpic}
 \end{center}
\caption{                     \label{fig:ingredients}
(Color online). a) The commutation relations $[H,G_{x}^{\nu}]$ guarantees that the gauge invariant subspace, i.e.~the trivial irreducible representation subspace for every lattice gauge subgroup, is dynamically decoupled from the rest of the Hilbert space. b) The nontrivial support of every lattice gauge generator is a single matter field site $\psi_x$ and all the gauge field links $U_{x,x+\mu_x}$ connected to it. c) Typical coupling Hamiltonian terms involve two matter sites $\psi_x$ and $\psi_{x+\mu_x}$ and the gauge boson connecting them $U_{x,x+\mu_x}$. d) In the QLM formulation, the gauge boson is split into a pair of rishons, linked together by a $U(1)$ symmetry constraint.}
\label{definition}
\end{figure}

As we mentioned before, there are two types of degrees of freedom in lattice gauge models, which we describe as finite-dimension quantum variables:
\begin{itemize}
\item Matter fields $\psi_{x}$ are located on the vertices of the lattice $x$. They are usually fermionic fields that describe the ``quarks'' of the model, $\{ \psi_{x}, \psi^{\dagger}_{y} \}=\delta_{x,y}$. But they can also be bosonic fields describing, for instance, the Higgs field. In non-abelian models, fermions $\psi^{a}_x$ carry color degrees of freedom $a$. For example, in $U(2)$ or $SU(2)$ models  $a \in \{ \uparrow, \downarrow \}$, in $U(3)$ or $SU(3)$ models $a \in \{  {\color{blue} b} , {\color{darkgreen} g} , {\color{red} r}  \}$

\item Gauge field $U^{ab}_{x,x+\mu_{x}}$ live on the links of the lattice $\langle x,x+\mu_{x} \rangle$. They are bosonic fields that describe the gauge bosons of the model. We use the quantum link formulation to
recast these fields as bilinear operators: $U^{ab}_{x,x+\mu_{x}} = c^{a \dagger}_{x+\mu_{x},-\mu_{x}} c^{b}_{x,+\mu_{x}}$, as sketched in figure \ref{fig:ingredients}, panel d.
As discussed in Refs.~\cite{Wiese2, Brower}, such bilinear decomposition is well-defined once a representation of the
symmetry group has been selected for the gauge boson $U^{ab}$.
In result of this formulation, every lattice link now hosts \emph{two} field modes, typically called `rishons' in usual terminology of quantum link models, and respectively labeled as $\langle x, +\mu_x \rangle$ and $\langle x+\mu_x, -\mu_x\rangle$.
The meaning of these auxiliary modes will become clear when we elaborate in detail some particular cases and models. Nontheless, we advance that they can be seen as a generalization of the Schwiger representation for the gauge field $U^{ab}$.
\end{itemize}

Such bilinear representation of the gauge fields
can be made either fermionic or bosonic, by setting the appropriate commutation relations for these operators $[c^{b}_{x,\mu_x},c^{a \dagger}_{y,\mu'_y}]_{\pm} =\delta_{a,b} \delta_{x,y} \delta_{\mu_x, \mu'_y}$. The statistics of the rishon fields $c^{a}_{x,\mu_x}$ is completely arbitrary, and does not change the statistics of the original gauge bosons $U^{ab}_{x,x+\mu_x}$, since the rishon operators $c_{x,\mu_x}^{a}$ always appear in pairs related to the same link.
Notice, however, that due to the hardcore nature of fermionic statistics,
fermionic rishons pose limits to the maximal number of rishons per link.
The total number of rishons $\eN_{x,x+\mu_x}= n_{x+\mu_{x},-\mu_{x}}  + n_{x,+\mu_{x}} = \sum_a ( c^{a \dagger}_{x+\mu_{x},-\mu_{x}} c^{a}_{x+\mu_{x},-\mu_{x}} + c^{a \dagger}_{x,+\mu_{x}} c^{a}_{x,+\mu_{x}} )$ on every link is a conserved quantity. This is due to the fact that the rishon degrees of freedom $c^{a}_{x,\mu_x}$ appear both in the gauge symmetry operators $G_{x}^{\nu}$ and in the Hamiltonian $H$ only via $U^{ab}_{x,x+\mu_x}$, and by construction $[\eN_{x,x+\mu_x},U^{ab}_{y,y+\mu_y}] =0$: from this follows that $[\eN_{x,x+\mu},G_y^{\nu}] = [\eN_{x,x+\mu}, H] = 0$. In other words, in the QLM formulation of lattice gauge theories, an additional, artificial local symmetry arises: the conservation law of the total rishons number on a given link, which is always $U(1)$ symmetry generated by $\eN_{x,x+\mu_x}$
(regardless of the symmetry group generated by $G^{\nu}_x$ which may as well be non-abelian).
There are different representations of the same symmetry depending on the number of rishons per link $\baN$ one selects. In any case, we restrict the Hilbert space to the `physical' states $|\varphi_{\text{phys}}\rangle$ which satisfy
$ \eN_{x,x+\mu_x}|\varphi_{\text{phys}}\rangle = |\varphi_{\text{phys}}\rangle \baN_{x,x+\mu_x}$. For simplicity, we will refer to this symmetry selection rule as \emph{link constraint},
as opposed to the \emph{gauge constraint} which is generated by $G_{x}^{\nu}$ instead (see next paragraph).
With little abuse of notation, in cases where the total number of rishons on a link $\baN_{x,x+\mu_x}$ is independent of the link itself (\ie homogeneous and isotropic QLM) we will sometimes omit the link label subscript.

\subsubsection{Local generators of the gauge symmetry, and gauge constraint (Gauss' law).}

The gauge symmetry is defined via the set of its generators $G^{\nu}_{x}$: they all commute with the Hamiltonian $[H,G^{\nu}_{x}]=0$, and have localized support. To properly characterize the generators $G^{\nu}_x$, it is convenient to define the elementary transformation on the gauge fields beforehand.
We will separately consider in the following the abelian and non-abelian parts: $U(1)$ and $SU(N)$.

\begin{itemize}
\item
{\bf Abelian $U(1)$:} here
the elementary transformation is generated by the difference of the rishon occupation numbers on the same link, \ie$E_{x,x+\mu_{x}} = \frac{1}{2} \left(n_{x+\mu_{x},-\mu_{x}}  - n_{x,+\mu_{x}} \right)$, which plays an equivalent role of the \emph{electric field} in quantum electrodynamics. Its action on the gauge field changes the field with a phase, 
\begin{equation}
\tilde{U}^{ab}_{x,x+\mu_{x}} =  e^{i\theta E_{x,x+\mu_{x}}} U^{ab}_{x,x+\mu_{x}} e^{- i\theta E_{x,x+\mu_{x}}} = e^{i\theta} U^{ab}_{x,x+\mu_{x}},
\end{equation}
or infinitesimally $\left[ E_{x,x+\mu_{x}}, U^{ab}_{x,x+\mu_{x}}\right]=U^{ab}_{x,x+\mu_{x}}$.

\item
{\bf Non-abelian $SU(N)$:} in this scenario, the corresponding non-abelian version of the electric field has a left component $L^{\nu}_{x,x+\mu_{x}} = \sum_{ab} c^{a\dagger}_{x,+\mu_{x}} \frac{\lambda^{\nu}_{ab}}{2} c^{b}_{x,+\mu_{x}}$ and a right component $R^{\nu}_{x,x+\mu_{x}} = \sum_{ab} c^{a\dagger}_{x+\mu_{x},-\mu_{x}} \frac{\lambda^{\nu}_{ab}}{2} c^{b}_{x+\mu_{x},-\mu_{x}}$ operators, depending if their action changes the bosonic gauge field $U^{ab}_{x,x+\mu_{x}}$ with a unitary $\Omega_{ak}$ acting on the left or on the right of the field
\begin{equation}
\begin{split}
\tilde{U}^{ab}_{x,x+\mu_{x}} =  e^{i \sum_{\nu} \theta^{\nu} L^{\nu}_{x,x+\mu_{x}}} U^{ab}_{x,x+\mu_{x}} e^{- i \sum_{\nu} \theta^{\nu} L^{\nu}_{x,x+\mu_{x}}} = \sum_{k} \Omega_{ak} U^{k b}_{x,x+\mu_{x}} \\
\bar{U}^{ab}_{x,x+\mu_{x}} =  e^{i \sum_{\nu} \theta^{\nu} R^{\nu}_{x,x+\mu_{x}}} U^{ab}_{x,x+\mu_{x}} e^{- i \sum_{\nu} \theta^{\nu} R^{\nu}_{x,x+\mu_{x}}} = \sum_{k} U^{a k}_{x,x+\mu_{x}} \Omega_{bk}^{*},
\end{split}
\end{equation}
or infinitesimally $\left[ L^{\nu}_{x,x+\mu_{x}} , U^{ab}_{x,x+\mu_{x}} \right] = - \sum_{k} \lambda^{\nu}_{ak} U^{kb}_{x,x+\mu_{x}}$ and $\left[ R^{\nu}_{x,x+\mu_{x}} , U^{ab}_{x,x+\mu_{x}} \right] =  \sum_{k}  U^{ak}_{x,x+\mu_{x}} \lambda^{\nu}_{kb}$, where $\lambda^{\nu}$ are the Hermitian generators of $SU(N)$ which obey $[\lambda^{\mu}, \lambda^{\nu}]=2i f_{\mu\nu\omega}\lambda^{\omega}$, with $f_{\mu\nu\omega}$ the structure constants of the $SU(N)$ algebra and $\text{Tr}\left( \lambda^{\mu} \lambda^{\nu} \right) =2 \delta^{\mu \nu}$.

\end{itemize}

Having properly defined the elementary transformations of the gauge fields, we can now easily introduce the complete gauge symmetry generators.

\begin{itemize}
\item  The local generator of the $U(1)$ part of the gauge model is defined by
\begin{equation}
\begin{split}
G_{x} &= \psi^{\dagger}_{x} \psi_{x} + \sum_{\mu_{x}} \left( E_{x-\mu_{x},x} - E_{x,x+\mu_{x}} \right)  \\
& = \psi^{\dagger}_{x} \psi_{x} + \sum_{\mu_{x}} \left( n_{x,-\mu_{x}} + n_{x,+\mu_{x}} \right) 
- \frac{ \bar{N}_{x-\mu_x,x}+ \bar{N}_{x,x+\mu_x}}{2},
\end{split}
\end{equation}
where in the first line, we have expressed the generator in terms of the electric field component, while in the second line, the generator is written in terms of matter and rishon fields. Thanks to the the QLM formulation, it is possible to write $G_{x}$ as an operator acting only on the QLM degrees of freedom on vertex $x$. Around every vertex, the gauge-invariance (or gauge-covariance) constraint is given by
\begin{equation}
\left [\psi^{\dagger}_{x} \psi_{x} + \sum_{\mu_{x}} \left( n_{x,-\mu_{x}} + n_{x,+\mu_{x}} \right)  \right] |\varphi_\text{phys}\rangle = |\varphi_\text{phys}\rangle \times \text{constant},
\end{equation}
which is equivalent to a Gauss' law, and the physical states $|\varphi_\text{phys}\rangle$ are those which satisfy it.

\item The generators for the non-abelian $SU(N)$ part of the gauge transformations fulfill the usual algebra $[G^{\mu}_{x}, G^{\nu}_{y}] = i \epsilon^{\mu \nu \omega} G^{\omega}_{x} \delta_{x,y}$ with
\begin{equation}
\begin{split}
G^{\nu}_{x} =&\sum_{ab} \psi^{a\dagger}_{x} \frac{\lambda^{\nu}_{ab}}{2} \psi^{b}_{x} + \sum_{\mu_{x}} \left( R^{\nu}_{x-\mu_{x},x}  + L^{\nu}_{x,x+\mu_{x}} \right) \\
=&\sum_{ab} \left[ \psi^{a\dagger}_{x} \frac{\lambda^{\nu}_{ab}}{2} \psi^{b}_{x} + \sum_{\mu_{x}} \left( c^{a\dagger}_{x,-\mu_{x}} \frac{\lambda^{\nu}_{ab}}{2} c^{b}_{x,-\mu_{x}}  + c^{a\dagger}_{x,+\mu_{x}} \frac{\lambda^{\nu}_{ab}}{2} c^{b}_{x,+\mu_{x}} \right)  \right]
\end{split}
\end{equation}

Again, in QLM formulation $G^{\nu}_{x}$ acts on matter and rishon fields belonging to lattice site $x$ only.
The gauge invariant subspace corresponds to the trivial irreducible representation subspace of the symmetry group generated by $G^{\nu}_{x}$, \ie the singlet subspace: $G^{\nu}_{x} | \varphi_\text{phys} \rangle =0$. This provides an extension to non-abelian gauge symmetries of the of the Gauss' law.

\end{itemize}

It is important to stress that every single element of the algebra $G^{\nu}_x$ is local as it acts nontrivially only on the degrees of freedom sharing the vertex $x$, as this will be the key ingredient for the computational speed-up.
Such gauge symmetry locality is sketched in Fig.~\ref{fig:ingredients}.b: every gauge generator acts only on one matter field site $\psi_x^{a}$ and $z$ gauge fields 'sites' $U^{ab}_{x,x+\mu_x}$ (or rishon sites $c^{a}_{x,\mu_x}$ in QLM formulation), with $z$ being the lattice coordination number, \ie all the quantum degrees of freedom sharing vertex $x$.

As a result, the total number of generators $G_x^{\nu}$ in the whole lattice gauge algebra scales extensively with the system size, a property which dramatically reduces the manifold dimension the system lives in, as we will see later on.

At the same time,
the combined gauge invariance constraints actig on a given vertex $x$,
can be assembled into a single linear mapping, which reads
\begin{equation} \label{eq:contractND}
|j_x\rangle_r = \sum_{s_{\psi}, \vec{s}_{\mu_x}} A^{[x] j}_{s_{\psi}, \vec{s}_{\mu_x}} |s_{\psi}, \vec{s}_{\mu_x} \rangle_x,
\end{equation}
once a canonical basis for the matter field $|s_{\psi}\rangle_x$, and one for each rishon field $|s_{\mu_x}\rangle_{x,\mu_x}$, have been chosen. This mapping defines
a `reduced' local basis $|j_x\rangle_r$ which spans exactly and solely the local gauge-invariant subspace. In the next sections, the states $|j_x\rangle_r$ will be adopted as logical, or computational, basis for all numerical purposes, and will be the starting platform for building a Tensor Network formulation upon.

\subsubsection{Gauge invariant dynamics}

The last element that has to be to defined is a gauge invariant model, its dynamics formulated via the Hamiltonian $H$. By construction, a gauge invariant Hamiltonian must commute with the local generators of the gauge symmetry
and those of the link symmetry in the QLM formulation, i.e.  $[H,G^{\nu}_{x}]=[H,\eN_{x,x+\mu_x}]=0$. Clearly, the class of Hamiltonians satisfying these requirements is still extremely wide. Here we will focus on short-range Hamiltonians that encompass the physics of typical lattice gauge models.

A pure gauge model, which embeds non-abelian gauge symmetry content, is given by the following Hamiltonian:
\begin{equation}
\begin{split}
H_{\text{pure}} = & H_{\text{electric}} + H_{\text{magn}} \\
= &\sum_{x, \mu_{x}} \left\{ g^{2}_{\text{abel}} \left( E_{x,x+\mu_{x}} \right)^{2} +g^{2}_{\text{non-ab}}
\sum_{\nu} \left[ \left( L^\nu_{x,x+\mu_{x}} \right)^{2} + \left( R^{\,\nu}_{x,x+\mu_{x}}\right)^{2} \right] \right\} \\
&- \frac{1}{g_{\text{magn}}^{2}} \sum_{x, \mu_{x}, \mu_{y}} \left[ \text{Tr} \left( U_{x,x+\mu_{x}} U_{x+\mu_{x},x+\mu_{x}+\mu_{y}} U_{x+\mu_{x}+\mu_{y},x+\mu_{y}} U_{x+\mu_{y},x} \right) + \text{h.c.} \right].
\end{split}
\end{equation}
The electric terms, respectively abelian and non-abelian, quantify the energy of the flux for the abelian or non-abelian part of the gauge group. While the first term encourages to have zero abelian electric flux on the link, the second favors singlets of rishons in the non-abelian color variables.
The magnetic term associates a positive energy density to every non-zero magnetic flux on every plaquette.
$g^{2}_{\text{abel}}$, $g^{2}_{\text{non-ab}}$ and $g_{\text{magn}}^{2}$ are the coupling constants for the abelian part of the electric field, non-abelian part and magnetic term, respectively.
A physically meaningful choice of constants is the one that recovers the Kogut-Susskind (KS) Hamiltonian~\cite{Kogut}:
that is $g^{2}_{\text{magn}} = 4a g^{2}$, and $g^{2}_{\text{abel}} = g^{2}_{\text{non-ab}}= \frac{ g^{2}}{2a}$,
where $a$ is the lattice spacing. Indeed, with this special choice of the couplings, one expects to recover the physics
of the usual $U(N)$ gauge theories in the continuum limit. Alternatively, one expects to approach the strong coupling
limit by setting $g_{\text{abel}} \simeq g_{\text{non-ab}} \gg 1/g_{\text{magn}}$.
It is important to remark that the previous quantum-link Hamiltonian satisfies a $U(N)$ gauge invariance by construction:
it is however possible to reduce the gauge symmetry into a pure $SU(N)$ by adding
artificial Hamiltonian terms which explicitly break the $U(1)$ part of the gauge symmetry, as proposed in Refs.~\cite{Wiese2, Brower}.

The coupling of the gauge fields with the matter fields is done with the lattice version of the ``minimal'' coupling, i.e. a hopping term of fermions mediated by the gauge field. Also, the mass term of the fermions is a gauge invariant term, hence,
\begin{equation} \label{eq:gaucoup}
H_{\text{coup}} = \sum_{x, \mu_{x}} J_{x,\mu_{x}} \left(  \psi^{\dagger}_{x} U_{x,x+\mu_{x}} \psi_{x+\mu_{x}} + \text{h.c.} \right) + \sum_{x} m_{x} \psi^{\dagger}_{x} \psi_{x}
\end{equation}
where we have defined site dependence hopping constants $J_{x,\mu_{x}}$ and mass term $m_{x}$, in case a specific distributions of signs, depending on the sites, is needed for a particular type of fermion introduced on the lattice. This type of minimal coupling is also sketched in Fig.~\ref{fig:ingredients}, panel c.

\subsubsection{Examples} \label{sec:exa1}

We have presented all the ingredients that are necessary to define a quantum link version of a lattice gauge theory, however for the sake of clarity and concreteness, we now present four particular examples: the simplest $(1+1)$ dimensional Quantum Link Model with the abelian $U(1)$ symmetry, the simplest $(1+1)$ dimensional Quantum Link Model with non-abelian $U(2)$ symmetry, and then, an application to two relevant models in condensed matter physics: quantum dimer \cite{dimer1, dimer2, dimer3} and spin ice \cite{spinice1, spinice2} models on the square lattice \cite{dimerice}.
~\\
\paragraph{\bf \emph{U(1)} Quantum Link Model} \label{U13states}

- The gauge invariant quantum Hamiltonian is given by
\begin{equation}
H= J \sum_{x} \left(  \psi^{\dagger}_{x} U_{x,x+1} \psi_{x+1} + \text{h.c.} \right) + g^{2} \sum_{x} \left( E_{x,x+1} \right)^{2} + m \sum_{x} \left( -1 \right)^{x} \psi^{\dagger}_{x} \psi_{x} 
\end{equation}
where the last term is a staggered chemical potential profile for the matter field, which is a spinless fermion field $\{\psi_{x}, \psi^{\dagger}_{y} \}=\delta_{x,y}$. Here $J$ is the strength of the matter-gauge field coupling, $g^{2}$ the electric-field energy density and $m$ the staggered mass. The gauge fields can be written in terms of rishons $U_{x,x+1}=c_{x,+} c^{\dagger}_{x+1,-}$, which are bosonic in nature $[ c_{x,a} , c_{y,b}^{\dagger}] = \delta_{x,y} \delta_{a,b}$.

The two independent local symmetries in this $U(1)$ Quantum Link Model are:
\begin{enumerate}
\item Constant number of rishons per link: $\eN_{x,x+1} |\varphi_{\text{phys}}\rangle = ( n_{x+1,-}  + n_{x,+} ) |\varphi_{\text{phys}}\rangle =  \baN \, |\varphi_{\text{phys}}\rangle$
\item Gauss' law on every vertex: $\left( \psi^{\dagger}_{x} \psi_{x} + n_{x,-} + n_{x,+} \right) |\varphi_{\text{phys}}\rangle = |\varphi_{\text{phys}}\rangle \left( \baN - \frac{1+ \left( -1 \right)^{x} }{2} \right)$
\end{enumerate}

The unusual factor $\left( \baN - \frac{1+ \left( -1 \right)^{x} }{2} \right)$ appears because we introduce $\psi_{x}$ spinless fermionic operators (matter fields with a staggered mass term $m$) usually denoted as staggered fermions~\cite{Kogut,Kogut2}. The vacuum of the staggered fermions is given by a quantum state at half-filling describing the Fermi-Dirac sea.

In what follows, we would like to understand in more detail two limits depending on the occupation $\baN$. Thus, we characterize the action of the gauge operators and electric field operators on a Hilbert space defined by the occupation of rishons $n_{x,+}$ and $n_{x+1,-}$ or equivalent by the total number of rishons on the link $\eN_{x,x+1}=\baN$ and the electric flux $E_{x,x+1}=\frac{n_{x+1,-} - n_{x,+}}{2}$, i.e., $|n_{+}, n_{-}\rangle = |\baN,E\rangle$, where we have omitted the labels of the link $\langle x, x+1 \rangle$:
\begin{itemize}
\item $\baN \gg 1$ (Wilson limit)~\cite{zohar2013}: Wilson formulation of compact $U(1)$ gauge theories starts with an infinite local dimensional Hilbert space defined with two conjugate variables: the electric field $E$ and an angle $\vartheta$, that fulfill the usual commutation relation of position and momentum $[E,\vartheta]=i$. Then, defining the link operator $U=e^{-i\vartheta}$, it is straightforward to check that $\left[ U, U^{\dagger}\right] = 0$, $\left[ E, U \right] = U$ or in an eigenstate basis of the electric field operator $U |E\rangle = |E+1\rangle$.

In $U(1)$ QLM for general occupation $\baN$, the link operator and the electric field fullfil $U | \baN,E\rangle = \sqrt{ \frac{\baN}{2} \left( \frac{\baN}{2} +1\right) - E \left( E +1 \right)} |\baN,E+1\rangle$ and $\left[ U, U^{\dagger}\right] = E$. In the limit $\baN\gg E$,
\begin{multline}
\frac{1}{\sqrt{ \frac{\baN}{2} \left( \frac{\baN}{2} +1\right)}  }U | \baN,E\rangle \to |\baN,E+1\rangle; ~ \, ~ \frac{1}{ \frac{\baN}{2} \left( \frac{\baN}{2} +1\right)  } \left[ U, U^{\dagger}\right] \to 0; \\
 \frac{1}{\sqrt{ \frac{\baN}{2} \left( \frac{\baN}{2} +1\right)}  }\left[ E, U \right] = \frac{1}{\sqrt{ \frac{\baN}{2} \left( \frac{\baN}{2} +1\right)}  }U
\end{multline}
which are the usual definition of the Wilson type lattice theories if we identify $\frac{1}{\sqrt{ \frac{\baN}{2} \left( \frac{\baN}{2} +1\right)}  }U$ with a unitary operator or parallel transporter of a $U(1)$ gauge model.
\item The other extreme limit is $\baN=1$: In this case there is only one rishon per link and the dimension of the gauge invariant Hilbert space around every vertex is three, having one empty mode and two occupied on the odd vertices and two empty modes and one occupied on the even ones.
\end{itemize}

\paragraph{\bf \emph{U(2)} Quantum Link Model}

- The generators of the $SU(2)$ gauge transformations fulfill the usual algebra $[G^{\mu}_{x}, G^{\nu}_{y}] = i \epsilon^{\mu \nu \omega} G^{\omega}_{x} \delta_{x,y}$ with
\begin{equation}
\begin{split}
G^{\nu}_{x} =&R^{\nu}_{x} + \sum_{a,b} \left(\psi^{a\dagger}_{x} \frac{\sigma^{\nu}_{ab}}{2} \psi^{b}_{x} \right) + L^{\nu}_{x} \\
=& \sum_{a,b} \left( c^{a\dagger}_{x,-} \frac{\sigma^{\nu}_{ab}}{2} c^{b}_{x,-} + \psi^{a\dagger}_x \frac{\sigma^{\nu}_{ab}}{2} \psi^{b}_x  + c^{a\dagger}_{x,+} \frac{\sigma^{\nu}_{ab}}{2} c^{b}_{x,+} \right)
\end{split}
\end{equation}
The gauge invariant subspace corresponds to a singlet of this operator, i.e. $G^{\nu}_{x} | \varphi_\text{phys} \rangle =0$.
A $U(2)$ gauge invariant Hamiltonian can be written as
\begin{equation}
\begin{split}
H=& \frac{1}{2} \sum_{x} \left( g^2_{\text{a}} \,E^{2}_{x} 
+ g^2_{\text{na}} \sum_{\nu} \left[ (R^{\nu}_{x})^2 + (L^{\nu}_{x})^2 \right]
\right)  + m \sum_{x,a} \left( -1 \right)^{x} \psi^{a \dagger}_{x} \psi^{a}_{x} \\
&+ t \sum_{x,a,b} \left[ \psi^{a \dagger}_{x} U^{ab}_{x,x+1} \psi^{b}_{x+1} + \text{h.c.} \right] 
\end{split}
\end{equation}
The $g^2_{\text{a}}$ and $g^2_{\text{na}}$ terms describe respectively the abelian and non-abelian electric field energy contributions, $m$ represents the staggered mass and $t$ the interaction between matter and gauge fields.
The non-abelian part of the gauge selection rule requires that the matter and rishon particles (both spin $\frac{1}{2}$)
on a vertex to form a color singlet, therefore they must be an even number.
Still, the possible combinations of total particle number on a vertex $n_{x, \psi} + n_{x,-} + n_{x,+}$ and
on a link $\baN_{x,x+1}$ are various.
A possibility, discussed here, is the configuration that includes the uniform half-filling matter state (one matter fermion per vertex).
In 1D, a simple way to achieve it is by setting $\baN = 1$, and $n_{x, \psi} + n_{x,-} + n_{x,+} = 2$.
The local gauge invariant basis is four dimensional: $\{ |\uparrow, \downarrow, 0 \rangle,~ |\uparrow, 0 , \downarrow \rangle,~ |0, \uparrow, \downarrow \rangle,~ |0, \phi, 0 \rangle\}$,
where $|\uparrow, \downarrow \rangle \equiv \frac{1}{\sqrt{2}}  \left( |\uparrow, \downarrow \rangle - | \downarrow , \uparrow \rangle \right)$, and $|\phi\rangle$ is the doubly-occupied site, with the two spin-$\frac{1}{2}$ particles forming a spin singlet.
Later on, we will consider the former scenario as an example, and also discuss a slightly more complex configuration (with fermionic rishons, $\baN = 2$ rishons per link, and $3 - (-1)^x$ particles on vertex $x$).

\begin{figure}
 \begin{center}
 \begin{overpic}[width = 360pt, unit=1pt]{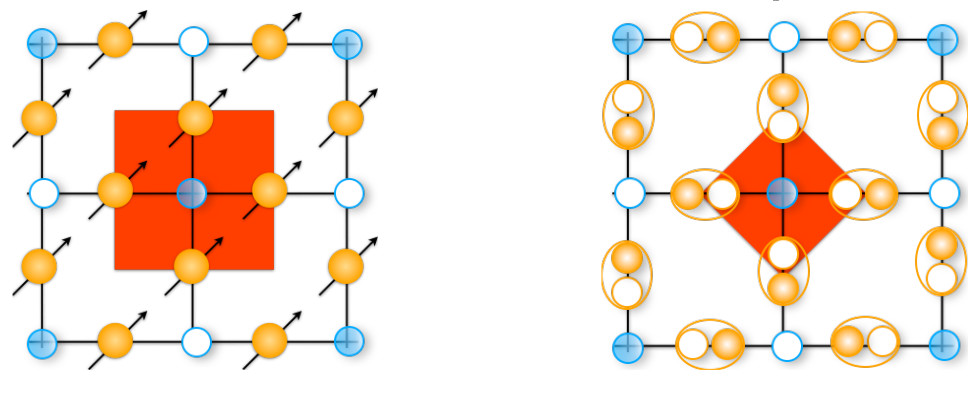}
  \put(-15, 126){\large a)}
  \put(200, 126){\large b)}
 \end{overpic}
 \end{center}\caption{           \label{fig:spinice}
 (Color online). Gauge generators supports in a) the standard formulation and in b) the quantum link formulation of lattice gauge theories of frustrated spin systems. 
 Blue circles represents sites of the lattice, orange ones the link degrees of freedom, i.e. the spins.
a) The red square on the lattice highlights the degrees of freedom on which the local gauge invariant generator acts: a site and the connected links. 
b) The red diamond shows the degrees of freedom on which the local gauge invariant generator acts for a QLM.
The original link
degree of freedom is split into two rishons, which are modeled by spinless fermions in this context.
}
\label{gauge}
\end{figure}
~\\
\paragraph{\bf Quantum dimer and spin ice models}
- In these models the matter field is fixed, and constitutes no quantum degree of freedom. The dynamics involves only gauge degrees on freedom, which are encoded in spins (hereafter we use spins-$\frac{1}{2}$ for simplicity) living on the links of a square lattice. The gauge symmetry generators are built upon one component of the Pauli matrices vector, say the third one $\sigma^{z}_{x,x+\mu}$. The spin-ice and dimer model share the same gauge symmetry generator, which reads
\begin{equation}
G_{x} = \sigma^{z}_{x,x+\mu_{x}} +  \sigma^{z}_{x,x+\mu_{y}}  + \sigma^{z}_{x-\mu_{x},x} + \sigma^{z}_{x-\mu_{y},x},
\end{equation}
however, in the two cases a different symmetry sector (irreducible subspace) is selected.
The QLM prescription splits the spin-$\frac{1}{2}$ in a pair of rishons, which are spinless
fermions in both cases: we thus rewrite 
$\sigma^{z}_{x,x+\mu} = \frac{1}{2} ( n_{x+\mu,-\mu} - n_{x,\mu} )$, obviously yielding
$[\sigma^{z}_{x,x+\mu}, \eN_{x,x+\mu}] = 0$. The link sector selected ($\baN = 1$,
\ie $\eN |\varphi_{\text{phys}}\rangle = |\varphi_{\text{phys}}\rangle$), recovers exactly a
two-level system on every link, as shown in Fig.~\ref{fig:spinice}.
 
In the quantum dimer model, the lattice is covered with dimer configurations on the links of the lattice. The dimer Hilbert space is characterized by the state $| | \rangle$ (on the vertical links) or $| - \rangle$ (on the horizontal links) with $\sigma^z_{x,x+\mu}=1/2$ if the link is occupied, otherwise the state is $|\,~ \rangle$ on the link and $\sigma^z_{x,x+\mu}=-1/2$. This model has been introduced  
to describe the presence of a Cooper pair or valence bond form by a pair of electrons on the nearest neighbor vertices (the dimers). The gauge constraint arises from the fact that every electron can pair only with one of the neighbor electrons, which results in  
the local conservation $\left( \sigma^z_{x,x+\mu_{x}} + \sigma^z_{x,x+\mu_{y}} +   \sigma^z_{x-\mu_{x},x} +  \sigma^z_{x-\mu_{y},x} \right) |\varphi_{\text{phys}}\rangle=-|\varphi_{\text{phys}}\rangle$. This gauge constraint reduces the Hilbert space from $2^{4}$ to just $4$ valid configurations around a vertex.

The quantum spin ice model is similar, but not identical. In this case the local gauge symmetry conservation originates from a strong antiferromagnetic Ising-type interaction between every pair of spins around a vertex:
\begin{equation}
H_{\text{Ising}} = \left( \sigma^z_{x,x+\mu_{x}} + \sigma^z_{x,x+\mu_{y}} +   \sigma^z_{x-\mu_{x},x} +  \sigma^z_{x-\mu_{y},x} \right)^{2}.
\end{equation}
Effectively this interaction projects the Hilbert space to the zero magnetization subspace $G_x |\varphi_{\text{phys}}\rangle = \left( \sigma^z_{x,x+\mu_{x}} + \sigma^z_{x,x+\mu_{y}} +   \sigma^z_{x-\mu_{x},x} +  \sigma^z_{x-\mu_{y},x} \right) |\varphi_{\text{phys}}\rangle = 0$. The local gauge invariant space is reduced to configurations with two spins $ |\!\uparrow\rangle$ and two spins $|\!\downarrow\rangle$ 
around a vertex, resulting in a local gauge vertex space dimension of $6$ instead of $2^4$.

\section{Matrix Product formulation of the QLM constraints}\label{sec:MPO}

In this section we embed the previous lattice gauge picture within the tensor network framework.
We first sketch a general technique, based on projected entangled pairs on the links,
which allows one to take operatively into account the QLM constraints defined previously,
 while reducing the computational space dimension, and thus the complexity of related algorithms.
The idea is to exploit the Gauge constraints to reduce the local space dimension, and at the same time combine all the link constraints into simple Projectors, which act directly upon the reduced space and, in 1D, are conveniently written in the Matrix Product Operator (MPO) formalism.

As we have seen in the previous examples, the gauge constraint and the link constraint in the QLM formulation 
result in a description of the system as composed by logical sites that groups a vertex of the original 
model and  the nearest neighbor interacting rishons sites.
Therefore, we can introduce a computational vertex site that is formed by the tensor product of a matter site and the rishons sites at that vertex, of compound dimension $D= d_\psi (d_c)^z$, where $d_\psi$ is matter local Hilbert space dimension, $z$ is the coordination number of the lattice, and $d_c$ is the local rishon space dimension (equal to $d_c = N+1$ in the abelian gauge case, larger otherwise). We show in the following that the gauge constraint can be solved by reducing the local site Hilbert space and that the remaining link constraints can be exactly written in a simple tensor structure that we can exploit to develop efficient 
implementations of numerical algorithms. 

Precisely, we restrict the local physical space to the trivial irreducible representation
subspace $|\varphi_{\text{phys}}\rangle_x$ of the local gauge symmetry group at vertex $x$, identified by
$G^{\nu}_x |\varphi_{\text{phys}}\rangle_x = 0$, with $G_x^{\nu}$ the group symmetry generators
we defined in the previous section.
Since gauge symmetries on different vertices commute, \ie
$[G^{\nu}_{x},G^{\nu'}_{x'\neq x}] = 0$, we can enforce the gauge requirement simultaneously on all vertices $x$.
Typical examples are the $SU(2)$ or $SO(3)$ gauge group cases, where the restricted states $|\varphi_{\text{phys}}\rangle$
are those vertex states which behave like a spin-0 under $G^{\nu}$.
Let now $P_{x}$ be the projector upon the physical space related to vertex $x$, and $|j_x\rangle_r$ an orthonormal basis 
for its range (which coincides with its support, since $P_{x} = P^2_{x} = P^{\dagger}_{x}$). The subscript $r$ indicates that we reduced the effective dimension to $d = \text{rnk}(P)$, since the rank of $P_{x}$ is always smaller than the original dimension $D$ of the combined degrees of freedom of vertex $x$, so that $d < D$. Then we have, for a one-dimensional QLM,
\begin{equation} \label{eq:contractor}
 |j_x\rangle_r = \sum_{s_{\psi}, s_-, s_+} A^{[x] j}_{s_-, s_{\psi}, s_+} |s_-, s_{\psi}, s_+\rangle_x,
\end{equation}
the generalization to any lattice and dimensionality is given by Eq.~\eqref{eq:contractND}.
The linear transformation of Eq.~\eqref{eq:contractor}
implements the map from the original $D$-dimensional basis to the $d$-dimensional basis of gauge constrained states,
and has a rectangular matrix representation $A$
with $j$ as row index and the combination of $s_{\psi}$, $s_-$ and $s_+$ as column
index (we dropped the vertex index $x$ for comfort of notation).
Since we chose an orthonormal reduced basis it follows that
\begin{multline}
 \delta_{j,j'} = \langle j| j' \rangle_r = \sum_{s_{\psi}, s_-, s_+} \sum_{s'_{\psi}, s'_-, s'_+}
 A^{[x] j}_{s_-, s_{\psi}, s_+} 
 A^{* [x] j'}_{s'_-, s'_{\psi}, s'_+}
 \langle s'_-, s'_{\psi}, s'_+ |s_-, s_{\psi}, s_+\rangle_x
 =\\=
 \sum_{s_{\psi}, s_-, s_+}
 A^{[x] j}_{s_-, s_{\psi}, s_+} 
 A^{* [x] j'}_{s_-, s_{\psi}, s_+},
\end{multline}
or, in matrix representation, $A A^{\dagger} = \Id_r$, \ie $A^{\dagger}$ is an isometry.
Similarly $A^{\dagger} A = P$, and thus $A P = A$.
The reduced basis $|j_x\rangle_r$ defines the local computational basis for any type of simulation on QLMs, since it generates
the full set of states fulfilling the gauge constraint.

In a Quantum Link Model formulation also the link constraint has to be satisfied simultaneously.
As previously stated,
the link symmetry group is always $U(1)$, and thus generated by a single operator per lattice link,
which reads $\eN_{x,x+1} = n_{x,+} + n_{x+1,-}$.
Here the operator $n_{x,\pm} = \sum_a c^{a \dagger}_{x,\pm} c^{a}_{x,\pm}$ counts the total number of rishons
in the mode $\langle x,\pm\rangle$, disregarding their color $a$.
By construction, the link group commutes
with the Hamiltonian,
\ie $[H, \eN_{x,x+1}] = 0$, as well as with the gauge group, \ie $[\eN_{x,x+1},G_{x'}^{\nu}] = 0$.
The link constraint requires that the number of rishons on the link $\langle x,x+1 \rangle$
is fixed to an integer number $\baN_{x,x+1}$, which means
$\eN_{x,x+1} |\varphi_{\text{phys}}\rangle =  \baN_{x,x+1} |\varphi_{\text{phys}}\rangle$.
The link constraint can be implemented by applying a projector
$Q_{x,x+1} = Q_{x,x+1}^2 = Q_{x,x+1}^{\dagger}$, which is diagonal as
every chosen rishon basis state $|s_{\mu}\rangle$ has a well defined occupation number
$n_{x,\pm} |s_\pm\rangle_{x,\pm} =  |s_\pm\rangle_{x,\pm} \bar{n}_{x,\pm}(s)$. In this case it reads
\begin{equation}
  Q_{x,x+1} = \sum_{s_-, s_+, q} B^{[x]}_{s_+, q} C^{[x+1]}_{q, s_-}\;
  |s_+\rangle\langle s_+|_x \otimes |s_-\rangle\langle s_-|_{x+1},
\end{equation}
where we split $Q_{x,x+1}$ according to its left-right Schmidt rank, resulting in
\begin{equation}
B^{[x]}_{s_+, q} = \delta_{\langle s_+ |n_{x,+}| s_+ \rangle,q}
\qquad \mbox{and} \qquad
C^{[x+1]}_{q, s_-} = \delta_{\baN_{x,x+1}-q,\langle s_- |n_{x+1,-}| s_- \rangle}.
\end{equation}
Of course, the fact that $[\eN_{x,x+1},G_{x'}^{\nu}] = 0$ implies that also $[Q_{x,x+1},P_{x'}] = 0$. Now, since all the $P_x$ act on mutually disjoint degrees of freedom for different $x$ (and so do the $A_x$ and the $Q_{x,x+1}$) we can define 
\begin{equation}
  \bar{P} = \bigotimes_{x = 1}^{L} P_x \quad, \quad
  \bar{A} = \bigotimes_{x = 1}^{L} A_x \quad \mbox{and} \quad
  \bar{Q} = \bigotimes_{x = 1}^{L - 1} Q_{x,x+1},
\end{equation}
representing the constraints combined over the whole lattice. Now we first enforce the link constraint, and then we contract the space onto the gauge-reduced basis. Basically, if we start from a generic, unconstrained many-body state $|\Psi\rangle$
we get
\begin{equation}
  \bar{A} \bar{Q} |\Psi\rangle = \bar{A} \bar{P} \bar{Q} |\Psi\rangle =
  \bar{A} \bar{Q} \bar{P} |\Psi\rangle = \bar{A} \bar{Q} \bar{A}^{\dagger} \bar{A} |\Psi\rangle =
  \bar{A} \bar{Q} \bar{A}^{\dagger} |\Psi\rangle_r = \bar{Q}_r |\Psi\rangle_r,
\end{equation}
where $|\Psi\rangle_r$ is now a generic many-body state in the gauge-reduced space, and
$\bar{Q}_r \equiv \bar{A} \bar{Q} \bar{A}^{\dagger}$ is the link constraint projector expressed in the reduced space.
Notice that $\bar{Q}_r$ is again a projector, since $\bar{Q}_r^2 =
\bar{A} \bar{Q} \bar{A}^{\dagger} \bar{A} \bar{Q} \bar{A}^{\dagger} = \bar{A} \bar{Q}^2 \bar{A}^{\dagger}
= \bar{A} \bar{Q} \bar{A}^{\dagger} = \bar{Q}_r$.
Moreover it is possible to write $\bar{Q}_r$ as follows:
\begin{equation} \label{eq:MPOform}
 \bar{Q}_r = \sum_{j_1 \ldots j_L}
 \sum_{j'_1 \ldots j'_L}
 \sum_{q_1 \ldots q_{L-1}}
  F^{[1]q_1}_{j_1,j'_1}\,
  F^{[2]q_1 q_2}_{j_2,j'_2}
  F^{[3]q_2 q_3}_{j_3,j'_3}
  \ldots
  F^{[L]q_{L-1}}_{j_L,j'_L}
  |j_1 \ldots j_L\rangle\langle j'_1 \ldots j'_L|_r
\end{equation}
where
\begin{equation} \label{eq:MPOmicro}
 F^{[x] q_{x-1}, q_x}_{j_x,j'_{x}} = \sum_{s_-, s_{\psi}, s_+}
 A^{[x] j_x}_{s_-, s_{\psi}, s_+}
 C^{[x]}_{q_{x-1}, s_-} B^{[x]}_{s_+, q_x}\,
 A^{* [x] j'_{x}}_{s_-, s_{\psi}, s_+}.
\end{equation}
Eq.~\eqref{eq:MPOform} is the Matrix Product Operator formulation of the projector $\bar{Q}_r$,
with the common index $q_\ell$ shared by two neighboring tensors, $F^{[\ell]}$ and $F^{[\ell+1]}$,
assuming
$m = \baN_{x,x+1} + 1$ distinct values (all integers from 0 to $\baN_{x,x+1}$).
Such integer $m$ is often referred to as \emph{bondlink dimension}, and it
has physical relevance in tensor networks formalism, since it relates to entanglement properties of the
state or operator described via tensor network ansatz~\cite{Ignazrev}.
For instance, in the DMRG, the entanglement entropy under
a left-right partition of the variational many-body state is bound by $\log m$.

Not surprisingly, the effective Hamiltonian expressed within the reduced space will preserve the link symmetry as it did in the original formulation. In fact, let $H_r = \bar{A} H \bar{A}^{\dagger}$ be the reduced Hamiltonian, then it holds
\begin{equation}
 [H_r, \bar{Q}_r] = [\bar{A} H \bar{A}^{\dagger}, \bar{A} \bar{Q} \bar{A}^{\dagger}]
 = \bar{A} [H, \bar{Q}] \bar{A}^{\dagger} = 0.
\end{equation}
In conclusion, in order to simulate the dynamics of a QLM, one can work completely in the reduced space and start the evolution in a quantum state of the form $\bar{Q}_r |\Psi_0 \rangle_r$, where $\bar{Q}_r$ enforces the link constraint. Then, the gauge-symmetric reduced Hamiltonian $H_r$ will preserve the link constraint since
\begin{equation}
|\Psi(t)\rangle_r = U_r(t) \bar{Q}_r |\Psi_0\rangle_r =
U_r(t) \bar{Q}_r^2 |\Psi_0\rangle_r = \bar{Q}_r U_r(t) \bar{Q}_r |\Psi_0\rangle_r = \bar{Q}_r |\Psi(t)\rangle_r,
\end{equation}
where $U_r(t) \equiv \exp(it H_r) = \bar{A} \exp(it H) \bar{A}^{\dagger}$. Moreover, it is possible to apply the projector $\bar{Q}_r$ at any time during state evolution, for instance to prevent the state from violating the
link constraint due to uncontrolled numerical errors.
As previously mentioned, the MPO formulation for the reduced link projector $\bar{Q}_r$ generalizes to any lattice and dimensionality
in a straightforward manner:
what one obtains is a Projected Entangled Pair Operator (PEPO), again with bondlink dimension bounded by
$\baN_{x,x+\mu_x}$.
\begin{figure}
 \begin{center}
 \begin{overpic}[width = \columnwidth, unit=1pt]{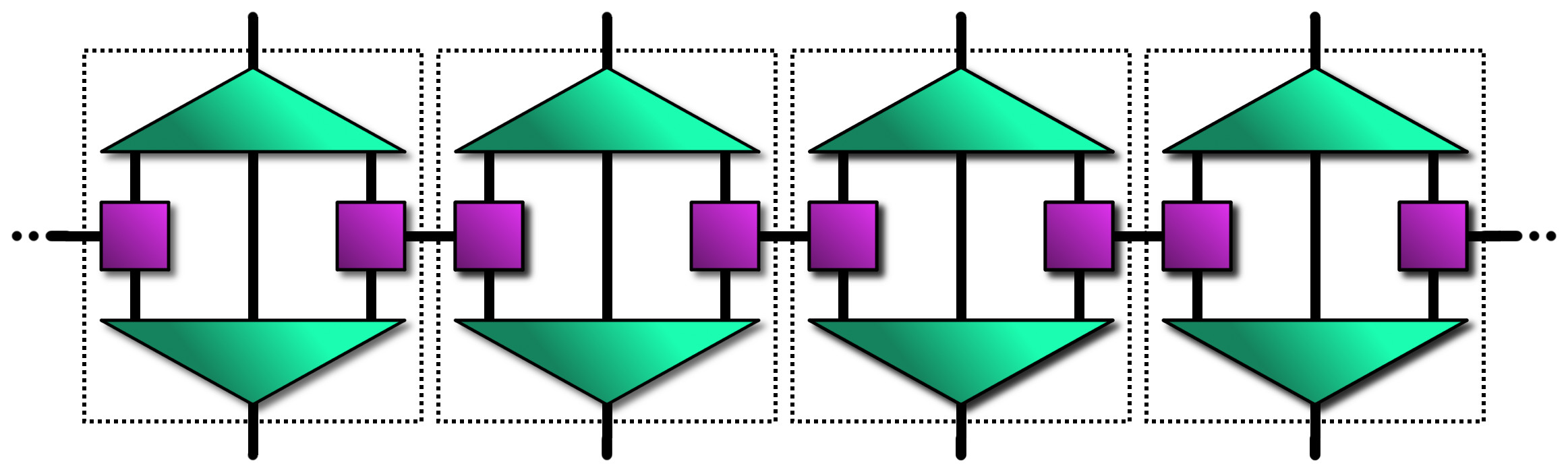}
  \put(66, 29){$A^{[x]}$}
  \put(62, 96){$A^{[x]\dagger}$}
  \put(29, 63){$C^{[x]}$}
  \put(96, 63){$B^{[x]}$}
  \put(26, 106){$F^{[x]}$}
  \put(162, 29){$A^{[x+1]}$}
  \put(156, 96){$A^{[x+1]\dagger}$}
  \put(132, 63){$\ldots$}
  \put(127, 106){$F^{[x+1]}$}
  \put(75, 48){$s_M$}
  \put(41, 48){$s_R$}
  \put(108, 48){$s_L$}
  \put(75, 1){$j_x$}
  \put(76, 127){$j'_x$}
 \end{overpic}
 \end{center}
\caption{ \label{fig:design} (color online)
Tensor network graphical diagram, representing the MPO formulation of the combined link constraint projector in the reduced basis space $\bar{Q}_r$. This picture corresponds to Eqs.~\eqref{eq:MPOform} and \eqref{eq:MPOmicro}.}
\end{figure}
\subsubsection{Canonical link-gauge basis}

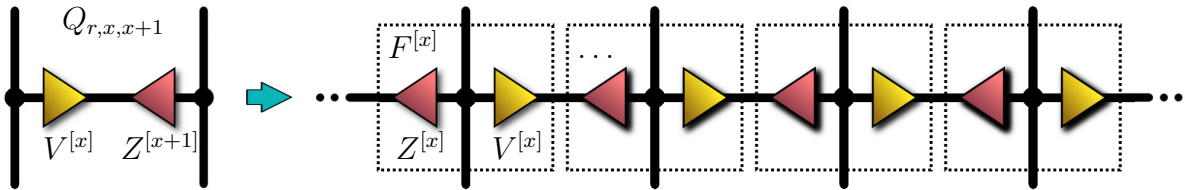
\begin{figure}
 \begin{center}
 \begin{overpic}[width = \columnwidth, unit=1pt]{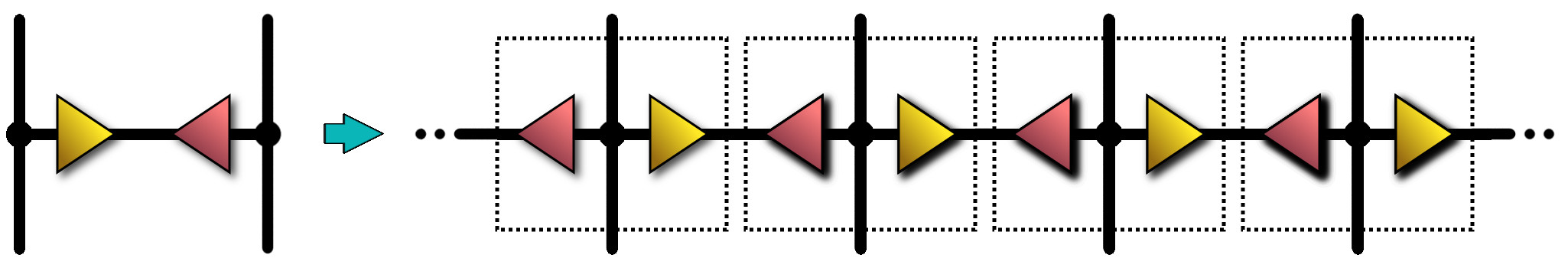}
  \put(23, 64){$Q_{r,x,x+1}$}
  \put(16, 15){$V^{[x]}$}
  \put(45, 15){$Z^{[x+1]}$}
  \put(145, 53){$F^{[x]}$}
  \put(148, 15){$Z^{[x]}$}
  \put(184, 15){$V^{[x]}$}
  \put(216, 53){$\ldots$}
 \end{overpic}
 \end{center}
\caption{ \label{fig:design2} (color online)
Tensor network graphical diagram of the $\bar{Q}_r$ in the canonical link-gauge basis. Left: the diagonal projector $Q_{r,x,x+1}$ decomposed according to Eq.~\eqref{eq:qudiagonal}. Right: simplified MPO representation of the combined link constraint in the reduced space $\bar{Q}_r$.
}
\end{figure}

As an additional remark, we show that introducing a particular basis $|j\rangle_r$ 
for the reduced space, the previous picture further simplifies: in the new basis $\bar{Q}_r$ reads as a diagonal operator while not increasing the previous MPO bond link dimension.
We start recalling that in the original QLM picture, the gauge generators $G^{\nu}_x$ conserve the number of rishons on their related links $n_{x,-}$ 
and $n_{x,+}$ 
separately, \ie
\begin{equation}\label{eq:allofthem}
 \left[ n_{x,-},G^{\nu}_{x'}\right] = \left[ n_{x,+},G^{\nu}_{x'}\right] = 0.
\end{equation}
This means that it exists a basis $|\psi_j\rangle_x$ in the space defined by $|s_-,s_\psi,s_+\rangle_x$ which diagonalizes simultaneously all the operators appearing in Eq.~\eqref{eq:allofthem}. Within this set, we identify those that satisfy the gauge constraint, and select them as the reduced basis $\left. |\psi_j\rangle_x \right|_{\text{phys}} \to |j_x\rangle_r$, precisely:
\begin{equation}
 n_{x,-}|j_x\rangle_r = |j_x\rangle_r \cdot \bar{n}_-(x,j)
 \quad \mbox{and} \quad
 n_{x,+}|j_x\rangle_r = |j_x\rangle_r \cdot \bar{n}_+(x,j).
\end{equation}
For obvious reasons, we refer to this special 
local basis choice as the canonical gauge-link basis. In this framework, the reduced link constraint projector reads
\begin{equation}\label{eq:qudiagonal}
 Q_{r,x,x+1} |j_x \, j'_{x+1}\rangle_r =
 |j_x \, j'_{x+1}\rangle_r
 \;\delta_{\bar{n}_{+}(x,j) + \bar{n}_{-}(x+1,j'),\baN_x}
 = |j_x \, j'_{x+1}\rangle_r \sum_{q = 0}^{\baN_{x,x+1}} V^{[x]}_{j_x,q} \cdot Z^{[x+1]}_{j'_{x+1},q}
\end{equation}
where simply we substituted $V^{[x]}_{j,q} = \delta_{\bar{n}_+(x,j),q}$ and
$Z^{[x+1]}_{q,j} = \delta_{\baN_{x,x+1}-q,\bar{n}_-(x+1,j)}$. Such simplified decomposition is sketched in Fig.~\ref{fig:design2}
(left panel). Notice that
$\baN_{x,x+1} + 1$ is exactly the Schmidt rank of the operator $Q_{r,x,x+1}$, so this decomposition is optimal in bondlink $m$ dimension.
Combining all the $Q_{r,x,x+1}$ together is straightforward now, since they are nearest-neighbor projectors diagonal in the reduced basis: doing so leads again to a MPO form of $\bar{Q}_r$ like Eq.~\eqref{eq:MPOform}, but with simpler tensor blocks:
\begin{equation} \label{eq:qubello}
 F^{[x]q_{x-1},q_x}_{j_x,j'_x} = \delta_{j_x, j'_x} \cdot Z^{[x]}_{q_{x-1},j_x} \cdot V^{[x]}_{j_x,q_{x}},
\end{equation}
as sketched in Fig.~\ref{fig:design2} (right panel). We know that this MPO representation is optimal in bondlink dimension $m$
because it uses the minimal bondlink to represent faithfully the Schmidt ranks of the matrices $Q_{r,x,x+1}$.
Such representation is extremely versatile: we will exploit it, for instance, to understand how QLM spaces dimensions (and thus computational costs) grow as a function of the total system size, in section \ref{sec:automata}.

\section{Fast link-constrained time-evolution scheme}  \label{sec:speedup}

As mentioned before, since the Hamiltonian commutes with every gauge or link symmetry in the original model, time-evolution
of the QLM dynamics should theoretically preserve all the constraints. Unfortunately, in numerical frameworks, systematic errors
are generated, and they may have dramatic, disruptive impact in conservation of symmetries
(if not addressed properly),
\eg in real-time evolution.
The imaginary-time evolution does not suffer from this issue: in fact, since local gauge symmetries can not
be spontaneously broken~\cite{Elitzur75}, convergence of the algorithm to the gauge-invariant ground state is guaranteed.
However, even in this scenario, addressing explicitly the gauge symmetry is computationally helpful:
setting non-gauge invariant states, which might be low-energy excitations, out of the variational picture,
can only speed-up the convergence rate to the ground state.
Moreover, the quasi-local constraints will allow us to speed up significantly the time-evolution algorithms
by performing all the linear algebraic operations in a computationally efficient block-wise fashion.

\subsection{\it Enforcing link constraints over time}

In this section we assume that we want to apply a (real or imaginary) time-evolution scheme of
a nearest-neighbor, time-independent QLM Hamiltonian $\bar{H}$ onto a many-body (unnormalized) mixed state $\rho$:
\begin{equation} \label{eq:realimag}
 \begin{aligned}
 \rho(t_0 + t) &= e^{i \bar{H} t} \rho(t) e^{-i \bar{H} t} \quad & \mbox{for real-time, or} \\
 \rho(\beta_0 + \beta) &= e^{- \bar{H} \beta/2} \rho(\beta_0) e^{- \bar{H} \beta/2} \qquad & \mbox{for imaginary-time}.
 \end{aligned}
\end{equation}
We also assume to have $\rho$ expressed variationally in a Matrix Product Density Operator (MPDO) formulation, i.e.
{
instead of numerically addressing $\rho$, we store the many-body operator $X$ such that $\rho = X X^{\dagger}$,
which always exists since $\rho > 0$, and we encode $X$ as a MPO. The time evolution can be then carried out directly on
$X$, because applying
\begin{equation} \begin{aligned}
 X(t_0 + t) &= e^{i \bar{H} t} X(t_0) \quad &\mbox{for real-time, or}\\
 X(\beta_0 + \beta) &= e^{- \bar{H} \beta/2} X(\beta_0) \qquad &\mbox{for imaginary-time},
 \end{aligned} \end{equation}
recovers exactly Eq.~\eqref{eq:realimag} via $\rho = X X^{\dagger}$ \cite{MPDO1}.}
Here we focus on nearest-neighbor Hamiltonians and thus it is convenient to evolve the state
by Time-Evolved Block Decimation (TEBD), a well-known procedure
in DMRG contexts based on Suzuki-Trotter (ST) decomposition of $\bar{H}$ into odd-even sites blocks and even-odd sites blocks~\cite{PrimoMPS3}.
More precisely,
\begin{multline} \label{eq:menost}
 \exp \left(\gamma \sum_x H^{[r]}_{x,x+1}\right) =
 \left( \bigotimes_x e^{c_1 \gamma H^{[r]}_{2x-1,2x} }\right)
 \left( \bigotimes_x e^{d_1 \gamma H^{[r]}_{2x,2x+1} }\right)
 \times \\ \times
 \left( \bigotimes_x e^{c_2 \gamma H^{[r]}_{2x-1,2x} }\right)
 \ldots
 \left( \bigotimes_x e^{d_p \gamma H^{[r]}_{2x,2x+1} }\right)
 + O(\gamma^p)
\end{multline}
where $p$ is known as ST-order and the coefficients $c_t$ and $d_t$ are calculated via Baker-Campbell-Hausdorff formula.
To enforce the link constraint, one might evolve the state via
$\bar{Q} \exp (\gamma \sum_x \bar{H}_{x,x+1})$. 
More in general one might want to apply
the link projector $\bar{Q}$ either before, or after the
evolution, or even both since $\bar{Q}^2 = \bar{Q}$; here we choose to apply it after the evolution step.
Here below we show that within the presented framework, this is straightforward and requires 
no additional computational cost.
We start showing that $\left[ Q_{r,x,x+1}, H_{r,x',x'+1} \right] = 0$, \ie even local Hamiltonian terms
commute with the link constraints. Indeed, 
\begin{multline}
\left[ Q_{r,x,x+1}, H_{r,x',x'+1} \right] = 
\left[ \bar{A} Q_{x,x+1} \bar{A}^{\dagger}, \bar{A} H_{x',x'+1}\bar{A}^{\dagger} \right]
=\\=
\bar{A} \left( Q_{x,x+1} \bar{P} H_{x',x'+1} - H_{x',x'+1} \bar{P} Q_{x,x+1} \right) \bar{A}^{\dagger}
= \\ =
\bar{A} \left( \bar{P} Q_{x,x+1} H_{x',x'+1} - H_{x',x'+1} Q_{x,x+1} \bar{P} \right) \bar{A}^{\dagger}
= \bar{A} \left[ Q_{x,x+1}, H_{x',x'+1} \right] \bar{A}^{\dagger},
\end{multline}
as $\bar{A} \bar{P} = \bar{A}$. In the original basis $Q_{x,x+1}$ and $H_{x',x'+1}$ act on common degrees of 
freedom only if $x = x'$, but also in this case the commutator is zero, since the local Hamiltonian term 
of Eq.~\eqref{eq:gaucoup}, respects the link symmetry on the inner bond.
Finally, 
\begin{multline} \label{eq:piust}
 \bar{Q} \exp \left(\gamma \sum_x \bar{H}_{x,x+1}\right) = 
 \left( \bigotimes_x M^{[r,1]}_{2x-1,2x} \right)
 \left( \bigotimes_x M^{[r,1]}_{2x,2x+1} \right)
 \times \\ \times
 \left( \bigotimes_x M^{[r,2]}_{2x-1,2x} \right)
 \ldots
 \left( \bigotimes_x M^{[r,p]}_{2x,2x+1} \right)
 + O(\gamma^p)
\end{multline}
where 
\begin{equation}
 M^{[r,\nu]}_{2x-1,2x} = Q_{r,2x-1,2x} \;\exp \left( c_{\nu} \gamma H^{[r]}_{2x-1,2x} \right), 
 \quad 
 M^{[r,\nu]}_{2x,2x+1} = Q_{r,2x,2x+1} \;\exp \left( d_{\nu} \gamma H^{[r]}_{2x,2x+1} \right). 
\end{equation}
This formulation ensures that the link symmetry is always protected without
increasing computational cost. We will see now that actually one can reduce such cost
by exploiting the constraint, and gain a significant speed-up of the algorithm.

\subsection{\it Link constraint computational speed-up}

Here we show that the link constraint formulation allows us to gain consistent advantage in both of the
two elementary operations on the MPDO architecture required to apply
$M^{[r,\nu]}_{x,x+1}$ on $X$ (remember that the many-body state is $\rho = X X^{\dagger}$), 
namely: 1.~the contraction and 2.~ the SVD-truncated separation. These two operations between multilinear tensors
are represented for comfort to the reader in Fig.~\ref{fig:speedup}.
Let us first recall that the MPDO design stores the `semi-state' $X$ in the form
\begin{equation} \label{eq:MPOX}
 X = \sum_{j_1 \ldots j_L = 1}^{d}
 \sum_{k_1 \ldots k_L = 1}^{b}
 \sum_{w_1 \ldots w_{L-1} = 1}^{m}
  X^{[1]w_1}_{j_1,k_1}\,
  X^{[2]w_1 w_2}_{j_2,k_2}
  X^{[3]w_2 w_3}_{j_3,k_3}
  \ldots
  X^{[L]w_{L-1}}_{j_L,k_L}\;
  |j_1 \ldots j_L\rangle_r \langle k_1 \ldots k_L|_b,
\end{equation}
where the correlation bondlink dimension $m$ and the bath bondlink dimension $b$ are both arbitrary (although $b \leq m^2 d$),
and determine the computational costs and the final numerical precision of the simulation.
On the other hand, we now order all possible triplets of labels $(j,j',q)_{x,x+1}$ so that the corresponding
state $|j_x,j'_{x+1}\rangle$, belongs to the support of $Q_{r,x,x+1}$, and $q$ is the `intermediate charge' of the pair
\ie $q = \bar{n}_{+}(x,j) = \baN_x - \bar{n}_{-}(x+1,j')$. All these triplets are collected into the set 
$\Omega_{x,x+1}$, and their number is $\chi = \#\Omega_{x,x+1}$, clearly with $\chi < d^2$. After these initial
remarks, we can study the two  operations separately:

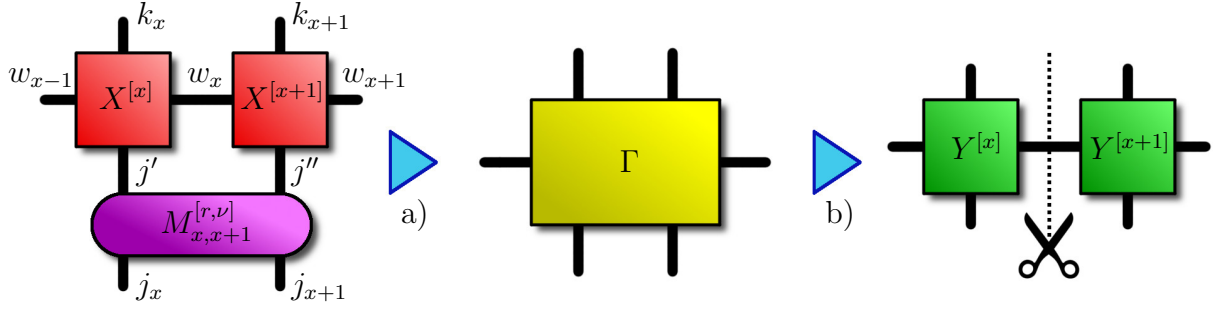
\begin{figure}
 \begin{center}
 \begin{overpic}[width = \columnwidth, unit=1pt]{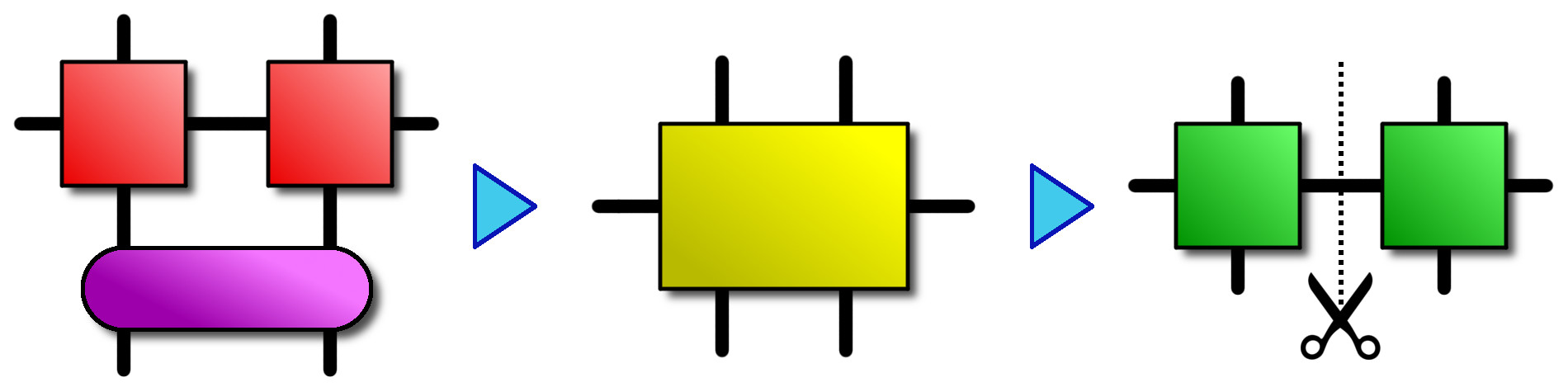}
  \put(49, 26){$M^{[r,\nu]}_{x,x+1}$}
  \put(26, 73){$X^{[x]}$}
  \put(79, 73){$X^{[x+1]}$}
  \put(40, 105){$k_x$}
  \put(98, 105){$k_{x+1}$}
  \put(40, 3){$j_x$}
  \put(98, 3){$j_{x+1}$}
  \put(40, 46){$j'$}
  \put(98, 46){$j''$}
  \put(-8, 83){$w_{x-1}$}
  \put(59, 83){$w_{x}$}
  \put(117, 83){$w_{x+1}$}
  \put(221, 49){$\Gamma$}
  \put(345, 54){$Y^{[x]}$}
  \put(397, 54){$Y^{[x+1]}$}
  \put(139, 30){a)}
  \put(298, 30){b)}
 \end{overpic}
 \end{center}
\caption{ \label{fig:speedup} (color online)
Pictorial Tensor-Network representation of the two basic operations which can be made faster by exploiting the link constraint:
a) Contraction; b) SVD-truncated separation.
}
\end{figure}

\begin{enumerate}
 \item {\bf Contraction - } The goal of this operation is to calculate entry-wise the tensor
  \begin{equation}
   \Gamma^{w_{x-1},k_x,j_x}_{w_{x+1},k_{x+1},j_{x+1}} = \sum_{w_x}^m \sum_{j',j''}^d
   \left( M^{[r,\nu]}_{x,x+1} \right)^{j',j''}_{j_x,j_{x+1}}
   X^{[x]w_{x-1} w_x}_{j',k_x}
   X^{[x+1]w_{x} w_{x+1}}_{j'',k_{x+1}},
  \end{equation}
whose cost normally scales (without considering fast matrix-multiplication schemes)
as $\sim d^2 b^2 m^3 + d^4 b^2 m^2$. The first term accounts for the cost of contracting the two $X$ tensors
together, the second term for assembling $M$. Exploiting the link symmetry in this procedure is
achieved by considering that both physical label pairs in input $(j',j'')$ and in output $(j_x,j_{x+1})$ must
satisfy the link constraint, i.e.~the triplets $(j',j'',q)$ and $(j_x,j_{x+1},q')$ must belong
to $\Omega_{x,x+1}$ for some $q$ and $q'$.
All the other pairs are identically zero both in input and output, and thus
need not to be considered in the computation. This remark reduces the computation as
\begin{equation}
 \mbox{cost} \sim d^2 b^2 m^3 + d^4 b^2 m^2 \qquad
 \mbox{reduced to} \longrightarrow \qquad
 \mbox{cost} \sim \chi\, b^2 m^3 + \chi^2 b^2 m^2
\end{equation}
\item {\bf SVD-truncated separation - } The second operation is needed to maintain the MPDO structure. Thus,
we split the $\Gamma$ tensor back into two blocks $Y$ via a Singular Value Decomposition (SVD), so that
\begin{equation}
   \Gamma^{w_{x-1},k_x,j_x}_{w_{x+1},k_{x+1},j_{x+1}}
   \simeq \sum_{w_x}^m
   Y^{[x]w_{x-1} w_x}_{j_x,k_x}
   Y^{[x+1]w_{x} w_{x+1}}_{j_{x+1},k_{x+1}}.
  \end{equation}
As usual, at every step one keeps
the correlation bond link dimension $m$ under control by discarding the smallest singular values.
Since the cost of a SVD for a $(s \times s)$-dimensional square matrix scales like $s^3$, the standard
cost for this operation is $\sim d^3 b^3 m^3$. This operation is speeded up exploiting the link constraint 
by observing that $\Gamma$ is shaped with an internal block structure.
In particular, an entire $(bm\times bm)$-dimensioned block
$(\Gamma^{j_x}_{j_{x+1}})^{w_{x-1},k_x}_{w_{x+1},k_{x+1}}$ is zero unless
$\Omega_{x,x+1}$ contains a triplet $(j_x, j_{x+1}, q)$.
Before performing the SVD, we reshuffle rows and columns of $\Gamma$ blockwise (this operation clearly preserves
the SVD decomposition). In particular, we reorder the rows so that $n_+(x,j_x)$ is monotonically increasing while descending the rows,
and we reorder the columns so that $n_-(x+1,j_{x+1})$ is monotonically decreasing while moving right in the columns.
Having done that, the resulting $\Gamma$ is block diagonal, with a number of blocks equal to the number
of intermediate charges $q$, usually (and always up to)
$\baN_x + 1$.
A single block, e.g. that related to intermediate charge $q$, has dimension
$(b \, m \, \xi(q) \times b \, m \, \xi(q))$ where $\xi(q)$ is the number of triplets in $\Omega_{x,x+1}$
containing intermediate charge $q$.
Finally, instead of performing SVD on the whole $\Gamma$ matrix, we perform separate SVD on each distinct $q$ block.
This approach reduces the computational costs as follows
\begin{equation}
 \mbox{cost} \sim d^3 b^3 m^3 \qquad
 \mbox{reduced to} \longrightarrow \qquad
 \mbox{cost} \sim b^3 m^3 \sum_q^{\baN_x} \xi^3(q).
\end{equation}
In the best case scenario, where the blocks have all roughly the same size $\xi = d/\baN_x$, the reduced cost
ultimately scales like $\sim (bmd/\baN_x)^3\baN_x \sim d^3 b^3 m^3 \baN^{-2}_x$, resulting in a net
$\baN^{-2}_x$ speed-up in the SVD procedure, which is often the computational bottleneck of the time-evolution algorithm.

\end{enumerate}

\section{Dimension of QLM spaces: the Cellular Automata} \label{sec:automata}

A fundamental issue that arises addressing numerically a quantum many-body problem is the amount of 
the computational resources required for the exact microscopic description scale with the total system size $\ell$.
This is a general problem which becomes even more relevant in the presence of symmetries: The constraints introduced by
the additional integrals of motion often reduce the Hilbert dimension scaling with $\ell$, up to the point where the
numerical complexity might change dramatically.
To understand how the one-dimensional QLM space dimension grows with the system size $\ell$ we work in the reduced basis,
while assuming an Open Boundary Conditions (OBC) setup so to proceed inductively by adding one site at a time, say from left to right.
We consider a QLM chain of length $\ell$; assume that we classified the `physical' states, which are of the form
$\bar{Q}_r(\ell)|\Psi\rangle_{r,\ell}$ according to the \emph{rightmost link charge} $q_\ell$. 
That is, we characterize a many-body orthogonal basis labeling the states via $|q_\ell,k\rangle_{r,\ell}$ so that
$n_{\ell,+} |q_\ell,k\rangle_{r,\ell} = q_\ell |q_\ell,k\rangle_{r,\ell}$, and the degeneracy label
$k\in\{1,D_q(\ell)\}$ spans within this charge sector. Clearly, the total Hilbert dimension is
given by $\bar{D}(\ell) = \sum_q D_q(\ell)$.
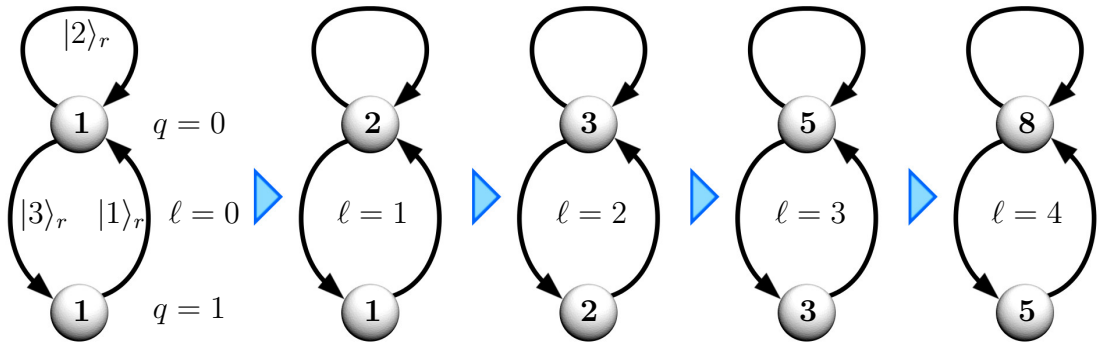
\begin{figure}
 \begin{center}
 \begin{overpic}[width = \columnwidth, unit=1pt]{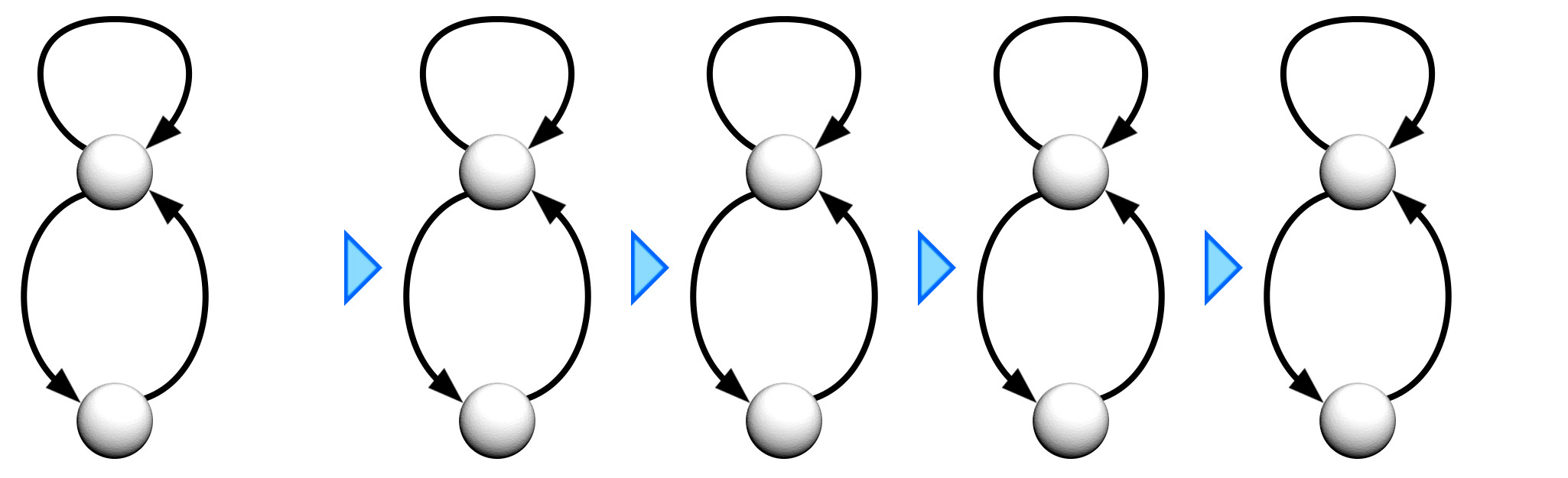}
  \put(60, 14){$q = 1$}
  \put(60, 84){$q = 0$}
  \put(10, 50){$|3\rangle_r$}
  \put(39, 50){$|1\rangle_r$}
  \put(26, 118){$|2\rangle_r$}
  \put(66, 50){$\ell = 0$}
  \put(30, 13){\bf 1}
  \put(30, 84){\bf 1}
  \put(129, 50){$\ell = 1$}
  \put(139, 13){\bf 1}
  \put(139, 84){\bf 2}
  \put(211, 50){$\ell = 2$}
  \put(220, 13){\bf 2}
  \put(220, 84){\bf 3}
  \put(293, 50){$\ell = 3$}
  \put(302, 13){\bf 3}
  \put(302, 84){\bf 5}
  \put(374, 50){$\ell = 4$}
  \put(384, 13){\bf 5}
  \put(384, 84){\bf 8}
 \end{overpic}
 \end{center}
\caption{ \label{fig:cellular} (color online)
Example of the Cellular Automata strategy to calculate the effective Hilbert dimension growth. This sketch corresponds to
the two-level abelian-$U(1)$ QLM, see section \ref{sec:exapmleCA}. The number written on node $q$ at
the stage $\ell$ of the automata corresponds to $D_q(\ell)$, i.e.~the total number of many-body states on $\ell$ sites,
satisfying all the gauge and link constraints, whose rightmost link charge is $q$.
}
\end{figure}

When adding a site to the previous picture, every (link-gauge) reduced basis state $|j_{\ell+1}\rangle_r$
connects only with those states $|q,k\rangle_{r,\ell}$ such that
$q= \baN_{\ell+1} -  \bar{n}_-(\ell+1,j)$. Moreover, every state of this form will have a well defined
rightmost link charge $q' = \bar{n}_+(\ell+1,j)$, and so they can be labeled again according to the rightmost link charge sectors.
Such inductive step will produce a new orthonormal complete basis of the type $|q',k'\rangle_{r,\ell+1}$. By construction
the new sector dimensions read
\begin{equation}
 D_{q'}(\ell + 1) = \sum_q \sum_j D_q(\ell) \; \delta_{q,\baN_{\ell+1} -  \bar{n}_-(\ell+1,j)} \;
 \delta_{\bar{n}_+(\ell+1,j),q'}
 = \sum_j D_{[\baN_{\ell+1} -  \bar{n}_-(\ell+1,j)]}(\ell)\;
 \delta_{\bar{n}_+(\ell+1,j),q'}.
\end{equation}
It can be useful for a clearer understanding, to encode
this recursive formula for calculating dimensions into a Cellular Automata.
The automata works according the following steps:
\begin{enumerate}
 \item Draw a node for each 
 link charge $q$ allowed
 \item Associate to every node the sector dimension $D_q(\ell)$. Starting step (zero sites): $D_q(0) = 1$ for every $q$.
 \item For every local reduced basis state $|j_{\ell+1}\rangle_r$ evaluate $q = \baN_{\ell+1} -  \bar{n}_-(\ell+1,j)$
  and $q' = \bar{n}_+(\ell+1,j)$. Then draw an arrow from node $q$ to node $q'$.
 \item The new sector dimensions $D_{q'}(\ell+1)$ are obtained by the sum, over all arrows that point to $q'$, of the dimension
  $D_q(\ell)$ of the node $q$ where that arrow starts from.
 \item Return to point 2 and iterate.
\end{enumerate}

In all the cases we considered, the dimension growth reduction due to the link symmetry is not stringent enough to make the scaling polynomial.
In fact, the scaling is still exponential $\bar{D}(\ell) \propto \alpha^{\ell}$
but the basis $\alpha$ is strictly smaller than the reduced local space dimension, \ie $\alpha < d$.
Moreover, we found $\alpha$ to be even smaller to the total number of allowed local matter states $|s_{\psi}\rangle$.
Before showing some examples on how the cellular automata works in practice, and what insight it can provide,
we wish to remind that the present scheme is meant only for one-dimensional quantum link models.
Indeed, higher-dimensionality lattices would require keeping track of the intermediate
charges for every open link when growing the lattice site-by-site, ultimately resulting in a more difficult treatment
which can not be trivially translated into a cellular automata paradigm.
This is nevertheless an interesting problem and it will constitute the focus for future research.

\subsubsection{Example: $U(1)$, spinless matter fermion, single rishon} \label{sec:exapmleCA}

This example corresponds to the QLM class introduce in paragraph \ref{U13states},
with the number of rishons per bond fixed to $N = 1$.
Here, both local matter and local gauge fields are two-level systems. The gauge constraint allows only
$d = 3$ states out of the $D = 8$ original ones to survive. Precisely,
written as $|s_-,s_{\psi},s_+\rangle$, they read
\begin{equation}
 \begin{aligned}
 &|1\rangle_r = |0,1,1\rangle_o \quad
 &|2\rangle_r = |1,0,1\rangle_o \quad
 &|3\rangle_r = |1,1,0\rangle_o \quad &\mbox{on odd sites}\\
 &|1\rangle_r = |1,0,0\rangle_e \quad
 &|2\rangle_r = |0,1,0\rangle_e \quad
 &|3\rangle_r = |0,0,1\rangle_e \quad &\mbox{on even sites.}
 \end{aligned}
\end{equation}
With the previous labeling of reduced basis states, the link constraint $Q_{r,x,x+1}$ becomes translationally-invariant
(\ie independent of $x$), and ultimately reads
\begin{equation}
 Q_{r,x,x+1} = |12 \rangle\langle 12|_r + |13 \rangle\langle 13|_r
 + |22 \rangle\langle 22|_r + |23 \rangle\langle 23|_r + |31 \rangle\langle 31|_r,
\end{equation}
with support dimension $\chi = 5$, or equivalently
\begin{equation}
 V_{j,q} = \left( \delta_{j,1} + \delta_{j,2} \right) \delta_{q,1} + \delta_{j,3} \,\delta_{q,2}
 \quad \mbox{and} \quad
 Z_{q,j} = \delta_{q,1} \left( \delta_{j,2} + \delta_{j,3} \right) + \delta_{q,2} \,\delta_{j,1},
\end{equation}
which requires a correlation bondlink $m = 2$, \ie we have two nodes in the cellular automata, respectively $q = 0$ and $q = 1$.
Regarding the automata connections we have:
State $|1\rangle_r$ connects $q = 1$ to the left, and $q = 0$ to the right, so it is an arrow from
$q = 1$ to $q = 0$. State $|2\rangle_r$ is an arrow from $q = 0$ to $q = 0$, while state $|3\rangle_r$ goes from $q = 0$ to $q = 1$.
A visual representation of this Cellular Automata is shown in Fig.~\ref{fig:cellular}. One can check immediately that the dimension
of this Hilbert space grows with $\ell$ exactly as the Fibonacci sequence: in fact $D_1(\ell) = D_0(\ell - 1)$ while
$D_0(\ell) = D_0(\ell-1) + D_1(\ell-1) = D_0(\ell-1) + D_0(\ell-2)$, and finally $\bar{D}(\ell) = D_0(\ell + 1)$.
In conclusion we have $\bar{D}(\ell) = (\varphi^{\ell + 3} - (1 - \varphi)^{\ell + 3})/\sqrt{5}$ with
$\varphi$ being the Golden Ratio $\varphi = (1 + \sqrt{5})/2$.
This tells us that for large sizes $\ell$, the Hilbert dimension grows exponentially $\bar{D}(\ell) \propto \alpha^\ell$, but instead of using as exponential basis the local space dimension $d = 3$
or the matter local dimension $d_{\psi} = 2$,
it is $\alpha = (1 + \sqrt{5})/2 \simeq 1.618$, \ie the scaling is fairly smoother.

\begin{figure}
 \begin{center}
 \begin{overpic}[width = \columnwidth, unit=1pt]{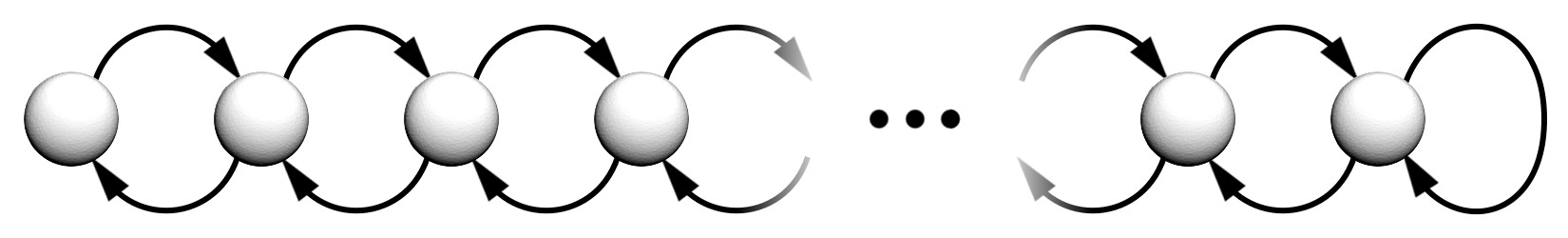}
 \end{overpic}
 \end{center}
\caption{ \label{fig:cell2}
 Sketch of the cellular automata for the $U(1)$ QLM with $N$ rishons on the link, where $N+1$ is the number of nodes in the picture.
 The rightmost node, the only one with a self-pointing arrow, is the one related to charge $q = \lfloor N/2 \rfloor$.
}
\end{figure}

\subsubsection{Example: $U(1)$, spinless matter fermion, multiple rishons} \label{sec:exaCA2}

\begin{figure}
 \begin{center}
 \begin{overpic}[width = 300pt, unit=1pt]{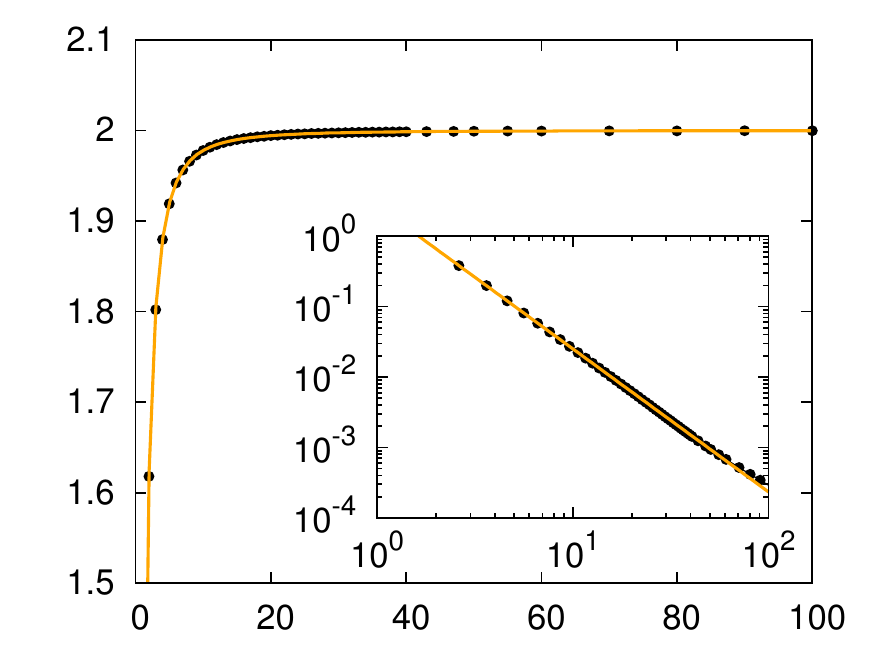}
  \put(155, -8){$\baN$}
  \put(0, 120){\large $\alpha$}
  \put(71, 103){$2 - \alpha$}
 \end{overpic}
 \end{center}
\caption{ \label{fig:scaloid} (Color online)
 Fitted exponential basis $\alpha$ of the Hilbert space dimension growth rate, as a function of the selected number of rishons
 per link $\baN$, in the $U(1)$ QLM scenario. Inset: distance of $\alpha$ from 2, plotted as a function of $\baN$,
 in double logarithmic scale.
}
\end{figure}

Again we explore the $U(1)$ QLM scenario, but this time we fix the number of rishons per link to be $\baN$, while the matter is
again a two-level quantum system. In this model there are $D = 2\baN^2$ local states available, however 
only $d = 2\baN + 1$ are allowed by the gauge constraint. We label them according to
\begin{equation}
 |2k+1\rangle_r = |k,1,\baN-k\rangle_o, \quad
 |2k\rangle_r = |k,0,\baN+1-k\rangle_o
\end{equation}
on odd sites, while on even sites
\begin{equation}
 |2k+1\rangle_r = |\baN-k,0,k\rangle_e, \quad
 |2k\rangle_r = |\baN-k,1,k-1\rangle_e,
\end{equation}
where $|j\rangle_r$ spans within $1 \leq j \leq 2\baN+1$. Like in the previous example $Q_{r,x,x+1}$ is homogeneous and reads
\begin{multline}
 Q_{r,x,x+1} = |2\baN+1, 1 \rangle\langle 2\baN+1, 1|_r
 + \sum_{k = 1}^{\baN}
 \left( |2k-1 \rangle\langle 2k-1|_r + |2k \rangle\langle 2k|_r \vphantom{\sum} \right)
 \otimes \\ \otimes
 \left( |2\baN-2k+3 \rangle\langle 2\baN-2k+3|_r + |2\baN-2k+2 \rangle\langle 2\baN-2k+2|_r \vphantom{\sum} \right)
\end{multline}
with support dimension $\chi = 4\baN + 1$. The allowed charges $q$ here go from $0$ to $\baN$, so we have
$\baN+1$ nodes in the cellular automata. The arrows are defined as follows: odd index states
and even index states connect the nodes as
\begin{equation}
 q_{\ell} = k \xrightarrow{| 2k+1 \rangle} q_{\ell+1} = \baN-k
 \quad \mbox{and} \quad
 q_{\ell} = k \xrightarrow{| 2k \rangle} q_{\ell+1} = \baN+1-k.
\end{equation}
After a reordering of all the nodes, the cellular automata appears as sketched in Fig.~\ref{fig:cell2}. We calculated numerically
the scaling of the total QLM Hilbert space dimension $\bar{D}(\ell)$ with the system size $\ell$, for several different rishon
number choices $\baN$. In every case considered, approximatively starting from sizes of $\ell \sim 10$,
$\bar{D}(\ell)$ matches an exponential scaling in $\ell$. We fitted the exponential basis $\alpha(\baN)$ 
$\bar{D}(\ell) \propto \alpha^{\ell}(\baN)$ in the interval $\ell \in [100, 1000]$. A smooth,
monotonic behavior of $\alpha$ as a function of $\baN$ is observed, and reported in Fig.~\ref{fig:scaloid}.
The calculated $\alpha$ values are never greater than 2, and saturate to 2 in a polynomial fashion with increasing numbers
of rishons per link $\baN$, \ie when approaching the Wilson limit, as shown in Fig.~\ref{fig:scaloid}, inset.
This study reveals that in proximity of the thermodynamical limit, the $U(1)$ quantum link model will never require more
computational resources or store more quantum information than an unconstrained spin-$\frac{1}{2}$ model,
which thus represents a comparative bound on the algebraic complexity of the quantum link model.
 This result seems to point in the direction that, in some cases, a $U(1)$ quantum link model (with spinless fermionic matter)
 could in principle be mapped into a spin-$\frac{1}{2}$ model with constraints, and these constraints
 vanish in the Wilson limit, where one should recover the corresponding unconstrained spin-$\frac{1}{2}$ model.
 Although our argument based on Hilbert dimension scaling does not provide any information whether such a mapping is
 actually possible for a given $U(1)$ gauge theory, some examples where the mapping exists are known.
 For instance, it was shown that the Schwinger model can be mapped to a long range interacting spin-$\frac{1}{2}$
model~\cite{Enrique13}.


\subsubsection{Example: $U(2)$, spin-$\frac{1}{2}$ fermions, single rishon} \label{sec:4statesU2ca}

In this example, which corresponds to the scenario we already introduced in section \ref{sec:exa1}, both matter and link fields
host spin-$\frac{1}{2}$ fermionic excitations. The $U(2)$ gauge constraint
fixes the available vertex states $|j\rangle_r$ to be both in a $SU(2)$ spin singlet
and in a defined $U(1)$ occupation number: here we consider the case $n_{x,\psi}+n_{x,-}+n_{x,+} = 2$.
Moreover, the link constraint fixes the number $\baN$ of rishons on a lattice bond, in this example to $\baN = 1$.
With the aforementioned choices, the reduced local space is four-dimensional:
\begin{equation}
\begin{aligned}
 |1\rangle_r =& \frac{1}{\sqrt{2}} \left( |0,\uparrow,\downarrow\rangle - |0,\downarrow,\uparrow\rangle \right),
 \qquad
 &|2\rangle_r =&\; \frac{1}{\sqrt{2}} \left( |\uparrow,0,\downarrow\rangle - |\downarrow,0,\uparrow\rangle \right),\\
 |3\rangle_r =& \frac{1}{\sqrt{2}} \left( |\uparrow,\downarrow,0\rangle - |\downarrow,\uparrow,0\rangle \right),
 \qquad
 &|4\rangle_r =&\; |0,\phi,0\rangle,
\end{aligned}
\end{equation}
where $|\uparrow\rangle \equiv c_{\uparrow}^{\dagger}|0\rangle$,
$|\downarrow\rangle \equiv c_{\downarrow}^{\dagger}|0\rangle$ and
$|\phi\rangle \equiv c_{\downarrow}^{\dagger} c_{\uparrow}^{\dagger} |0\rangle$. The reduced link projector
$Q_{r,x,x+1}$ has support dimension $\chi = 8$ and reads
\begin{equation}
 Q_{r,x,x+1} = \left( |1 \rangle\langle 1| + |2 \rangle\langle 2| \vphantom{\sum}\right) \otimes
 \left( |1 \rangle\langle 1| + |4 \rangle\langle 4| \vphantom{\sum}\right) + 
 \left( |3 \rangle\langle 3| + |4 \rangle\langle 4| \vphantom{\sum}\right) \otimes
 \left( |2 \rangle\langle 2| + |3 \rangle\langle 3| \vphantom{\sum}\right).
\end{equation}
\begin{figure}
 \begin{center}
  \begin{minipage}[t]{0.25\textwidth}
   \begin{overpic}[width = 60pt, unit=1pt]{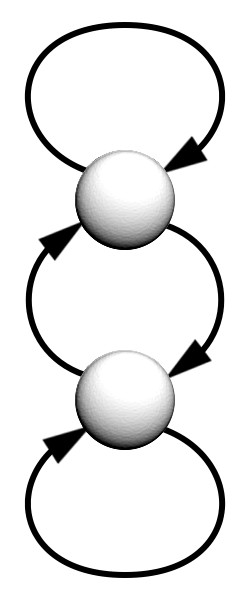}
    \put(-25, 150){a)}
    \put(-16, 94){$q = 0$}
    \put(-16, 45){$q = 1$}
   \end{overpic}
  \end{minipage}
  \begin{minipage}[t]{0.29\textwidth}
    \begin{overpic}[width = 70pt, unit=1pt]{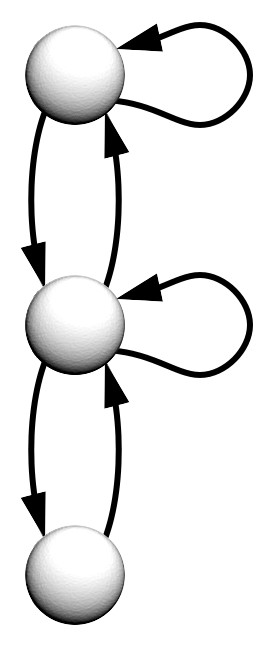}
     \put(-45, 150){b)}
     \put(-25, 15){$q = 0$}
     \put(-25, 79){$q = 2$}
     \put(-25, 143){$q = 1$}
    \end{overpic}
  \end{minipage}
  \begin{minipage}[t]{0.43\textwidth}
    \begin{overpic}[width =\textwidth, unit=1pt]{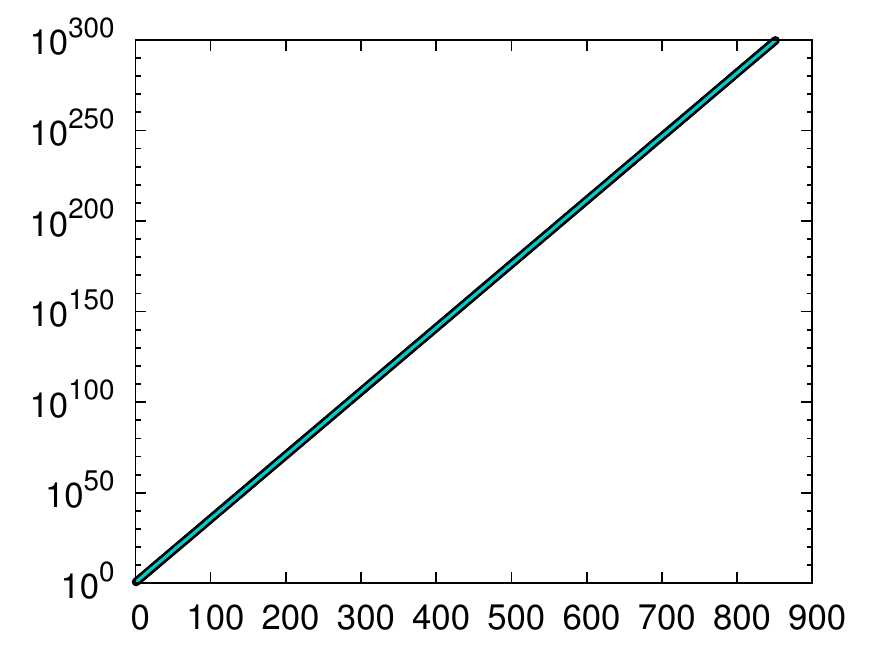}
     \put(-24, 150){c)}
     \put(-20, 70){$\bar{D}(\ell)$}
     \put(104, -10){$\ell$}
    \end{overpic}
  \end{minipage}
\end{center}
 \caption{ \label{fig:multibox} (Color online)
 Cellular automata scheme for the $U(2)$ QLM with spin-$\frac{1}{2}$ fermions and single rishon per link (a) and .
 for the double rishon per link (b) scenario. c) Scaling of the total Hilbert space dimension $\bar{D}(\ell)$
 as a function of the 1D chain length $\ell$, in the $U(2)$ double rishon case, evaluated numerically (black dots). The cyan
 line is an exponential fit, revealing the exponential basis $\alpha \simeq 2.2469796$.
}
\end{figure}
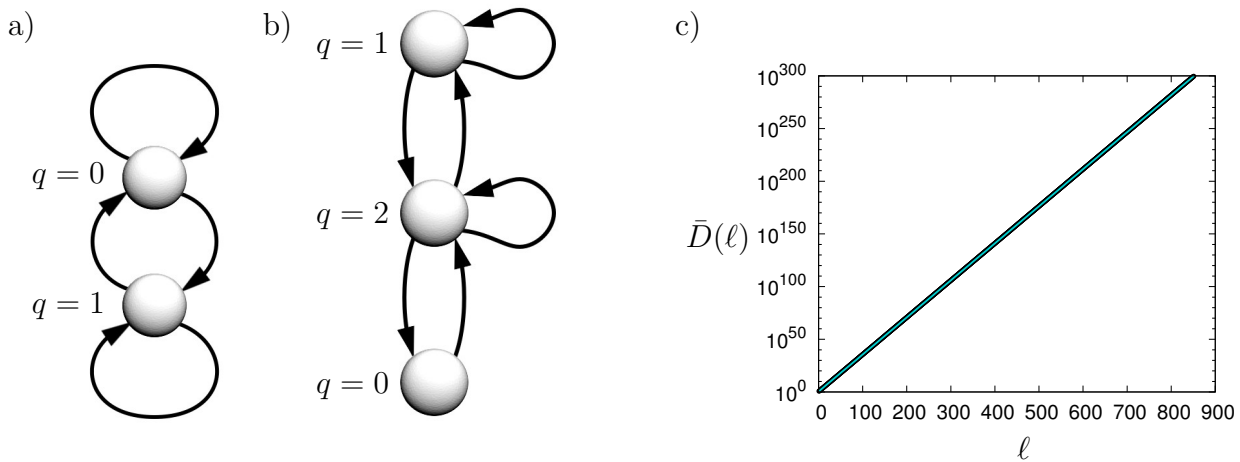
The corresponding cellular automata appears as shown in Fig.~\ref{fig:multibox}, left panel. The Hilbert space dimension scaling in this example is exactly an exponential, precisely it reads $\bar{D}(\ell) = 2^{\ell + 1}$, so that, ultimately, $\alpha = 2$.
{
This again hints to the fact that a mapping of this class of models to a spin-$\frac{1}{2}$ system might exist, even though --to the best of our knowledge--  it is not known.
}


\subsubsection{Example: $U(2)$, spin-$\frac{1}{2}$ fermions, double rishon} \label{sec:6statesU2ca}

The last scenario we discuss explicitly is again the $U(2)$ case with spin-$\frac{1}{2}$ particles,
but now we set $\baN = 2$ rishons on every link. The effective link constraint in the reduced formulation become
translationally invariant when we start from a staggered original fermion filling in the vertices: the choice that
describes more rich physics is given by 
$n_{x,\psi}+n_{x,-}+n_{x,+} = 3 - (-1)^{x}$, which results in $d = 6$ reduced states.
Indeed, on odd sites we have
\begin{equation}
 \begin{aligned}
 |1\rangle_r &= |\phi,\phi,0\rangle_o,
 \quad
 &|2\rangle_r &=\; \frac{1}{\sqrt{2}} \left( |\phi,\uparrow,\downarrow\rangle_o - |\phi,\downarrow,\uparrow\rangle_o \right),
 & \\
 |3\rangle_r &=\; \frac{1}{\sqrt{2}} \left( |\uparrow,\phi,\downarrow\rangle_o - |\downarrow,\phi,\uparrow\rangle_o \right),
 \quad
 &|4\rangle_r &=\; \frac{1}{\sqrt{2}} \left( |\uparrow,\downarrow,\phi\rangle_o - |\downarrow,\uparrow,\phi\rangle_o \right),
 & \\
 |5\rangle_r &=\; |\phi,0,\phi\rangle_o,
 \quad
 &|6\rangle_r &=\; |0,\phi,\phi\rangle_o,& \\
 \end{aligned}
\end{equation}
while on even sites
\begin{equation}
 \begin{aligned}
 |1\rangle_r &= |0,0,\phi\rangle_e,
 \quad
 &|2\rangle_r &=\; \frac{1}{\sqrt{2}} \left( |0,\uparrow,\downarrow\rangle_e - |0,\downarrow,\uparrow\rangle_e \right),
 \\
 |3\rangle_r &=\; \frac{1}{\sqrt{2}} \left( |\uparrow,0,\downarrow\rangle_e - |\downarrow,0,\uparrow\rangle_e \right),
 \quad
 &|4\rangle_r &=\; \frac{1}{\sqrt{2}} \left( |\uparrow,\downarrow,0\rangle_e - |\downarrow,\uparrow,0\rangle_e \right),
 \\
 |5\rangle_r &=\; |0,\phi,0\rangle_e,
 \quad
 &|6\rangle_r &=\; |\phi,0,0\rangle_e.\\
 \end{aligned}
\end{equation}
As there are three possible occupations on the link degree of freedom, the automata has three charges $q$.
The intermediate charge $q = 0$ connects $|1\rangle_r$ (to the left) to $|6\rangle_r$ (to the right),
charge $q = 1$ connects $\{|2\rangle_r, |3\rangle_r \}$ to $\{|3\rangle_r, |4\rangle_r \}$, and finally
$q = 2$ connects $\{|4\rangle_r, |5\rangle_r, |6\rangle_r \}$ to $\{|1\rangle_r, |2\rangle_r, |5\rangle_r \}$,
for a total reduced link projector support of dimension $\chi = 14 < d^2 = 36$.

The cellular automata for this setup is shown in Fig.~\ref{fig:multibox}, middle panel.
Using the automata mechanism, we numerically calculated the effective Hilbert space dimensions $\bar{D}(\ell)$ for
system sizes up to $\ell \sim 850$. Once again, an asymptotically exponential scaling $\bar{D}(\ell) \propto \alpha^{\ell}$
is detected: In
Fig.~\ref{fig:multibox}, right panel, we show how the exponential curve (cyan line) fits the data points,
(which have been enlarged on purpose not to be hidden by the fit curve).
The exponential basis we estimated from the fit is $\alpha \simeq 2.2470$.

\section{Conclusions} \label{sec:conc}

In this work we have merged the quantum link formalism with the tensor network framework and showed that combined they allow  
to efficiently describe equilibrium and out-of-equilibrium properties of lattice gauge theories in the Hamiltonian formulation. 
We showed how to combine efficiently gauge constraints and link constraints, in a matrix product operator 
formalism in 1D, which can be straightforwardly generalized to a projected entangled pair formalism in higher dimensions.
This paradigm is instrumental to merge time-evolution schemes, native to tensor network architectures,
with gauge-invariance constraints ultimately leading to a symmetry protected dynamics algorithm. 
Moreover, the local symmetries can be furthermore exploited to obtain a substantial enhancement in the algorithm performance.
Finally, we adopted the tensor network picture and developed a cellular automata formalism to compute the scaling of the gauge-invariant subspace 
of the quantum link models, and thus the effective complexity of the model. This analysis might be useful to estimate the computational complexity  
of a simulation of a given model and to guide the search for mappings from the original model to simplified ones.  

The framework introduced here will pave the way to the study of extremely interesting lattice gauge problems,
ranging from high energy physics in low dimensions up to topological condensed matter models. Indeed, 
global symmetries (e.g. conserved particle number or total magnetization) might be combined with 
this approach to achieve even higher performances and finite temperature or open system
dynamics~\cite{MPDO1, MPDO2, MPDODaniel},
richer tensor structures~\cite{Montavidal} and optimally controlled dynamics might be studied in the
future~\cite{CRAB1, CRAB2}.

~\\

ACKNOWLEDGEMENTS - Authors acknowledge support from EU through SIQS, ERC-St Grant ColdSIM (No. 307688), EOARD, 
UdS via Labex NIE and IdEX, and the German Research Foundation (DFG) via the SFB/TRR21. 
Authors thank M. Dalmonte and G. Pupillo for stimulating discussions.


\end{document}